\documentclass{article}

\usepackage{arxiv}

\usepackage[utf8]{inputenc} 
\usepackage[T1]{fontenc}    
\usepackage{hyperref}       
\usepackage{url}            
\usepackage{booktabs}       
\usepackage{amsfonts}       
\usepackage{nicefrac}       
\usepackage{microtype}      
\usepackage{lipsum}

\usepackage{graphicx}
\usepackage{epstopdf}
\usepackage{subfig}
\usepackage[justification=centering]{caption}
\hypersetup{
     colorlinks=true,
     linkcolor=blue,
     filecolor=blue,
     citecolor = blue,      
     urlcolor=cyan,
     }
     
\usepackage[nameinlink,capitalise]{cleveref}
\usepackage{xcolor}
\usepackage{colortbl}
\usepackage{amsthm}

\theoremstyle{plain}

\theoremstyle{definition}

\theoremstyle{remark}

\usepackage{filecontents}
\usepackage{multirow,fixltx2e}
\title{Dynamic System Identification of Underwater Vehicles Using Multi-Output Gaussian Processes}

\author{
  Wilmer Ariza Ramirez \\
  Australian Maritime College\\
  University of Tasmania\\
  Newnham, Tasmania, Australia\\
  \texttt{wilmer.arizaramirez@utas.edu.au} \\
   \And
  J. Kocijan \\
  Jožef Stefan Institute\\
  Jamova cesta 39, SI-1000\\
  Ljubljana, Slovenia\\
  School of Engineering and Management\\
  University of Nova Gorica\\
  Glavni trg 8, SI-5271 Vipava, Slovenia\\
  \texttt{jus.kocijan@ijs.si} \\
  \And
  Zhi Leong \\
  Australian Maritime College\\
  University of Tasmania\\
  Newnham, Tasmania, Australia\\
  \texttt{Zhi.Leong@utas.edu.au} \\
  \And
  Hung Nguyen \\
  Australian Maritime College\\
  University of Tasmania\\
  Newnham, Tasmania, Australia\\
  \texttt{H.D.Nguyen@utas.edu.au} \\
  \And
  Shantha Gamini Jayasinghe \\
  Australian Maritime College\\
  University of Tasmania\\
  Newnham, Tasmania, Australia\\
  \texttt{Shantha.Jayasinghe@utas.edu.au} \\
}

\begin{document}
\maketitle

\begin{abstract}
Non-parametric system identification with Gaussian Processes for underwater vehicles is explored in this research with the purpose of modelling autonomous underwater vehicle (AUV) dynamics with low amount of data. Multi-output Gaussian processes and its aptitude to model the dynamic system of an underactuated AUV without losing the relationships between tied outputs is used. The simulation of a first-principles model of a Remus 100 AUV is employed to capture data for the training and validation of the multi-output Gaussian processes. The metric and required procedure to carry out multi-output Gaussian processes for AUV with 6 degrees of freedom (DoF) is also shown in this paper. Multi-output Gaussian processes are compared with the popular technique of recurrent neural network show that Multi-output Gaussian processes manage to surpass RNN for non-parametric dynamic system identification in underwater vehicles with highly coupled DoF with the added benefit of providing a measurement of confidence. 
\end{abstract}

\keywords{Dependent Gaussian processes\and Dynamic system identification\and Multi-output Gaussian processes\and Non-parametric identification\and Autonomous underwater vehicles}

\section{Introduction}
Dynamic modelling of unmanned underwater vehicles (UUVs) has been a subject of interest among researchers since the early days of underwater exploration. Nowadays, UUVs are extensively employed in research, industry and military applications. Modelling of autonomous underwater vehicles (AUVs) is an important step for design of mission, control and navigation systems. Thus, accurate modelling and adaptability of such systems is an important issue due to such extensive applications. The most common methodologies use a mathematical model which is derived from Newtonian-Lagrange mechanics. This mathematical model is composed of a series of coefficients that need to be calculated to obtain an accurate model. Difference between the obtained model and the reality is usually treated in the literature as noise and in most cases, is modelled as a Gaussian distribution. A Gaussian distribution can be extended to the calculation of an approximation to a real model of a vehicle with higher exactitude and adaptability than a mathematical model\cite{RN81}.

Over the years, multiple methods for the calculation of coefficients of underwater vehicles have been proposed. \ One way to obtain the hydrodynamic coefficients is to perform a series of captive model tests such as rotating arm and planar motion mechanics \cite{RN145,RN144,RN122}. Another common technique for the hydrodynamic coefficient calculation is the use of computational fluid dynamics (CFD) \cite{RN147}. However, for the successful application, CFD still requires verification of results with experiments \cite{RN149}.

Nevertheless, research has probed the variability of mathematical models for AUVs as the vehicle operates in proximity to objects \cite{RN150}, near surface \cite{RN151} and most commercial available underwater vehicles with a modular architecture involving variable geometric and mass. Furthermore, in certain applications, the precision of some coefficients is require to be within 5\% of accuracy \cite{RN152}. Therefore, the variability of coefficients and the high precision required make it cumbersome or even impossible to acquire an exact analytical system model based on physical rules. 

Another procedure to obtain coefficients from a model of an underwater vehicle is the use of observers. Common observers applied to obtain the hydrodynamic coefficients of AUVs from measured data are least-squares \cite{RN153,RN154}, nonlinear Kalman filters such as extended (EKF) and unscented Kalman filter (UKF) \cite{RN155}. The EKF requires the linearization at each time step for the approximation of non-linearities which can be difficult to regulate and implement. A method to overcome this is the use of UKF which applies the unscented transform over a set of methodically chosen samples to model the system nonlinearity \cite{RN156}. Other common methodologies are frequency domain identification \cite{RN45}), neural networks (NN) \cite{RN158} and support vector machines (SVM) \cite{RN159}. The latter two methodologies are machine-learning algorithms and are more commonly used in online learning of the coefficients and also provide system adaptability. The adaptation of mathematical model has inherent defects such as the dependency of initial values, small quantity of coefficients to be updated, ill-conditioned matrix and drift. 

Machine learning algorithms are not limited to the calculation of hydrodynamic coefficients as they can learn to behave as part of the system or the complete system. Multiple applications have taken advantage of this ability and \ used NN \cite{RN160} and SVM \cite{RN78} to learn the damping model for the system which is placed in parallel to a well-known partial mathematical model. Other applications have used pure machine learning algorithms to identify a complete underwater vehicle as a black-box model with the use of nonlinear autoregressive model with exogenous (NARX) architecture. \cite{RN161} have used multiple architectures of NN for the regression of an AUV model in a NARX architecture and used the learned model for model predictive control. Their study shows that recurrent NN (RNN) provides higher faithfulness to the plant.

Recently, \cite{RN165} compared different machine learning algorithms for the system regression of underwater vehicles, i.e. NN, SVM, Gaussian Process Regression (GPR) and Kernel Ridge regression (KRR). \ Their results show that the machine learning algorithms could model an AUV from onboard sensor data in comparison to a least squares approach. \ Nevertheless, in their study, a structure for dynamic system identification has not been employed and each degree of freedom was treated as separate element. This can be problematic in AUVs as the outputs are strongly coupled. In the specific case of modelling with Gaussian Processes (GPs) \cite{RN81}, research shows that the dynamic regression of a system with GPs can produce better results than other methodologies. The most common methodology for Multi-Input-Multi-Output systems is to model each DoF as a separate system \cite{RN123}. More advance methodologies for Dynamic system identification have been proposed in \cite{RN124} and \cite{RN125} and specific methodologies are introduced for the identification of multi-output GPs based on the use of variation of dependent GPs.

GPs is a well-established methodology in fields such as geostatistics, where the \ method is called `kriging' \cite{krige1951statistical}. In GPs-based system identification, the model is built over input-output data and a covariance function is used to characterize the vehicle behaviour. The advantage of GPs is their ability to work with small quantities of data and noisy data. The predicted results consist of a mean and variance value which can be used for other purposes as well such as control, navigation and model based fault detection as it contains a measure of confidence. 

Multi-output GPs are a special case of GPs with the capability to model the nonlinear behaviour and coupling among outputs of a multi-output system \cite{RN172}, both which are important for AUV dynamics. In this study, a non-parametric dynamic system identification with Multi-Output GPs architecture employed by the authors for ships \cite{RN198} is extended to AUVs. The output from the algorithm will be a predictive value and a measure of confidence of the predictive value. The present implementation was made over data obtained from a non-conventional test with variable frequency of a nonlinear simulation model of a REMUS 100 AUV. Multiple sample times and data length were tested to find the best metric that can describe an AUV. A RNN was employed as a comparison to measure the effectiveness of the proposed method.
\section[Nonlinear Dynamic AUV Model]{Nonlinear Dynamic AUV Model}
\begin{figure}[!htb]
\centering
\includegraphics[scale=.4]{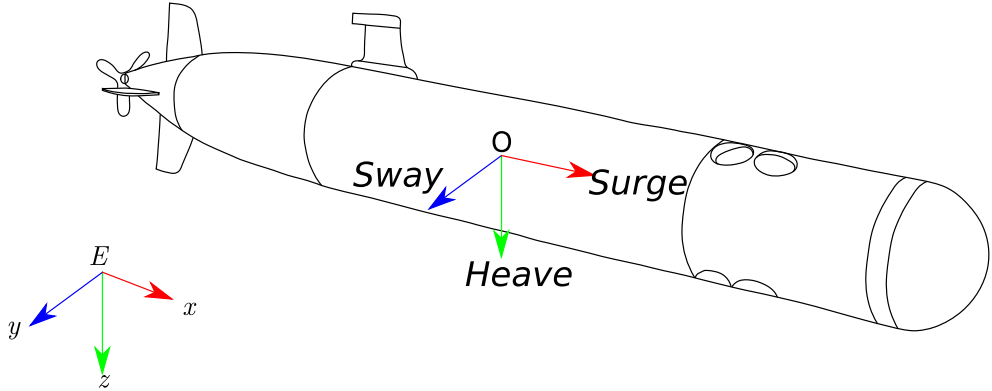}
\caption{AUV different reference frames, vehicle frame is equal to centre of buoyancy.}
\label{fig:figure-1}
\end{figure}

In \cite{RN76} it was shown that the nonlinear dynamic equations of motion of an underwater vehicle can be expressed in vector notation defined by a state vector composed by the vector $v$ of velocities on the body frame of the form ${\left[ {u,v,w,p,q,r} \right]^T}$ and the vector $\eta $ of position in the Earth fixed frame(\Cref{fig:figure-1}) of the form  ${\left[ {\xi ,\eta ,\zeta ,\phi ,\theta ,\psi } \right]^T}$ such that
\begin{equation}\label{eq3.1}
{\bf{M\dot v + C}}\left( {\bf{v}} \right){\bf{v + D}}\left( {\bf{v}} \right){\bf{v + g}}\left( {\bf{\eta }} \right){\bf{ = \tau }}
\end{equation}
with the kinematic equation

\begin{equation}
\label{eq3.2}
{\bf{\dot \eta  = J}}\left( {\bf{\eta }} \right){\bf{v}}
\end{equation}

where

${\bf{\eta }}$ \ \ position and orientation of the vehicle in Earth-fixed frame,

${\bf{v}}$ \ \ linear and angular vehicle velocity in body fixed frame,

${\bf{\dot v}}$ \ \ linear and angular vehicle acceleration in body fixed frame,

${\bf{M}}$ \ \ matrix of inertial terms,

${\bf{C}}\left( {\bf{v}} \right)$ \ \ matrix of Coriolis and centripetal terms,

${\bf{D}}\left( {\bf{v}} \right)$ \ \ matrix consisting of damping or drag terms

${\bf{g}}\left( {\bf{\eta }} \right)$ \ \ \ vector of restoring forces and moments due to gravity and buoyancy

${\bf{\tau }}$ \ \ vector of control and external forces

${\bf{J}}\left( {\bf{\eta }} \right)$ \ \ rotation matrix that converts velocity in a body fixed frame  $v$ \ to an Earth fixed frame velocity  $\dot \eta $ .

\Cref{eq3.1} can be expanded into a more general equation of motion as has been shown in \cite{RN178,gertler1967standard}.The result of the expansion will be a system of six equation with 73 hydrodynamic coefficients. However, for a complete model the control surfaces must be modelled. In a general case, the resulting forces and moments of a control surface (thrusters and fins) can be expressed as \cite{RN76}
\begin{equation}\label{eq3.3}
\begin{array}{l}
{F_{prop}} =  - {K_{fprop}}\left| n \right|n\\
{M_{prop}} =  - {K_{mprop}}\left| n \right|n
\end{array}
\end{equation}
\begin{equation}\label{eq3.4}
\begin{array}{l}
{L_{fin}} = {K_{\left. L \right|{\delta _{{\rm{fin}}}}}}{\delta _{{\rm{fin}}}}v_e^2\\
{M_{fin}} = {K_{\left. M \right|{\delta _{{\rm{fin}}}}}}{\delta _{{\rm{fin}}}}v_e^2
\end{array}
\end{equation}

A more accurate thruster model can be found \ in \cite{kim2006accurate} with the inclusion of the motor model and fluid dynamics. However, in this study, a more conservative model is used. Details of the Remus 100 AUV model used in this study are given in Section III.
\section{Dynamic Identification with Multi-output GPs}
GPs can be defined as a generalization of a multivariate Gaussian distribution. A multivariate Gaussian distribution is defined by it's mean and a covariance matrix. In the case of GPs is a distribution over functions rather a distribution over vectors. GPs is one of the methods based on kernel functions where the kernel function calculates the relationship between an input and an output point, and generates the covariance between them. The covariance determines how strongly linked (correlated) these two points are. In the case of multi-output GPs this is extended by the convolution of kernels to add not only the relationship between an input and an output but also the relationship between the outputs. The kernel is the key ingredient for the calculation of the covariance matrix that correlates inputs and outputs of training data \cite{RN171}.

The design of the algorithm for multi-output \ system identification with GPs is based on the previous work of Alvarez and Lawrence \cite{alvarez2009sparse} and \cite{RN81}. The dynamic identification problem can be defined as the search for relationship between a vector formed by delayed samples from the inputs ${\bf{u}}(k - 1)$  and outputs  ${\bf{y}}(k - 1)$ and the future output values. The relationship can be expressed by the equation:
\begin{equation}\label{eq3.5}
{\bf{y}}(k + 1) = f\left( {{\bf{x}}(k),\Theta } \right) + {\bf{v}}(k)
\end{equation}
where  $f\left( {{\bf{x}}(k),\Theta } \right)$ is a function that maps the sample data vector  ${\bf{x}}(k)$ to the output space based on the hyperparameters  $\Theta $;  ${\bf{v}}(k)$ accounts for the noise and error in the prediction of output  ${\bf{y}}(k)$. In the case of dynamic system identification the discrete time variable  $(k)$ is presented as an embedded element in the regression process as it is accounted in the delayed samples. 

A requirement for dynamic system identification of nonlinear systems is the selection of a nonlinear model structure such as nonlinear autoregressive model with exogenous input (NARX), nonlinear autoregressive (NAR), nonlinear output-error (NOE), nonlinear finite-impulse response (NFIR) and other structures. From all the possible structures, the simpler and most popular structure to implement is NARX as it only requires measurements of system output/s and input/s. In the case of an AUV, NARX is the most practical configuration since the measuring points are restricted to the available sensors \cite{RN81}. 
\subsection[Multi{}-output GPs]{Multi-output GPs}
The non-linear dynamic system of an AUV (\cref{eq3.1}) shows the level of coupling between the Newton-Lagrange equations of an AUV. The nonlinearity and coupling between outputs can be better represented by a multi-output GPs. Multi-output GPs presented here are based on the work of \cite{alvarez2009sparse}. Multi-output GPs are founded in the regression of data using convolving white noise process with a smoothing kernel function \cite{higdon2002space}. This was later introduced by \cite{boyle2005dependent} to the machine learning community by assuming multiple latent process defined over a space ${\Re ^q}$ . The dependency between two outputs is the model with a common latent process and their independency with a latent function, which does not interact with other outputs. If a set of functions $\left\{ {{f_q}\left( {\bf{x}} \right)} \right\}_{q = 1}^Q$ is considered, where  $Q$ is the Output Dimension for a  $N$ number of data points, where each function is expressed as the convolution between a smoothing kernel  $\left\{ {{k_q}\left( {\bf{x}} \right)} \right\}_{q = 1}^Q$ and a latent function  $u({\bf{z}})$ ,
\begin{equation}\label{eq3.6}
{f_q}({\bf{x}}) = \int_{ - \infty }^\infty  {{k_q}} \left( {{\bf{x}} - {\bf{z}}} \right)u\left( {\bf{z}} \right)dz
\end{equation}
This equation can be generalized for more than one latent function $\left\{ {{u_r}\left( {\bf{x}} \right)} \right\}_{r = 1}^R$ and includes a corruption function (noise) independent to each of the outputs \[{w_q}({\bf{x}})\] , to obtain
\begin{equation}\label{eq3.7}
\begin{array}{l}
{{\bf{y}}_q}\left( x \right) = {f_q}\left( {\bf{x}} \right) + {w_q}\left( {\bf{x}} \right)\\
{{\bf{y}}_q}\left( x \right) = \sum\limits_{r = 1}^R {\int_{ - \infty }^\infty  {{k_{qr}}\left( {{\bf{x}} - {\bf{z}}} \right)} } {u_r}\left( {\bf{z}} \right)d{\bf{z}} + {w_q}\left( {\bf{x}} \right)
\end{array}
\end{equation}
The covariance between two different functions  ${{\bf{y}}_q}\left( {\bf{x}} \right)$and  ${y_s}({\bf{x}}')$ is:
\begin{equation}\label{eq3.8}
\begin{array}{c}
{\mathop{\rm cov}} \left[ {{{\bf{y}}_q}\left( {\bf{x}} \right),{{\bf{y}}_s}({\bf{x'}})} \right] = {\mathop{\rm cov}} \left[ {{f_q}\left( {\bf{x}} \right),{f_s}({\bf{x'}})} \right]\\
 + {\mathop{\rm cov}} \left[ {{w_q}\left( {\bf{x}} \right),{w_s}({\bf{x'}})} \right]{\delta _{qs}}
\end{array}
\end{equation}
where
\begin{equation}\label{eq3.9}
\begin{array}{c}
{\mathop{\rm cov}} \left[ {{f_q}\left( {\bf{x}} \right),{f_s}({\bf{x'}})} \right] = \sum\limits_{r = 1}^R {\sum\limits_{p = 1}^R {\int_{ - \infty }^\infty  {{k_{qr}}({\bf{x}} - {\bf{z}})} } } \\
\int_{ - \infty }^\infty  {{k_{sp}}({\bf{x'}} - {\bf{z'}})} {\mathop{\rm cov}} \left[ {{u_r}\left( {\bf{z}} \right),{u_p}({\bf{z'}})} \right]dz'dz
\end{array}
\end{equation}
If it is assumed that  ${u_r}\left( {\bf{z}} \right)$ is an independent white noise ${\mathop{\rm cov}} \left[ {{u_r}\left( {\bf{z}} \right),{u_p}({\bf{z'}})} \right] = \sigma _{ur}^2{\delta _{rp}}{\delta _{z,z'}}$ will become:
\begin{equation}\label{eq3.10}
{\mathop{\rm cov}} \left[ {{f_q}\left( {\bf{x}} \right),{f_s}({\bf{x'}})} \right] = \sum\limits_{r = 1}^R {\sigma _{ur}^2\int_{ - \infty }^\infty  {{k_{qr}}({\bf{x}} - {\bf{z}}){k_{sp}}({\bf{x' - z'}})} d{\bf{z}}}
\end{equation}
The mean  ${\bf{\mathord{\buildrel{\lower3pt\hbox{$\scriptscriptstyle\frown$}} 
\over y'} }}$ with variance  ${{\bf{\sigma }}_{{\bf{\mathord{\buildrel{\lower3pt\hbox{$\scriptscriptstyle\frown$}} 
\over y'} }}}}$ of a predictive distribution at the point ${\bf{x'}}$ given the hyperparameters  ${\bf{\Theta }}$ can be defined as
\begin{equation}\label{eq3.11}
{\bf{\mathord{\buildrel{\lower3pt\hbox{$\scriptscriptstyle\frown$}} 
\over y'}  = k(x',x)k(x,x}}{{\bf{)}}^{{\bf{ - 1}}}}{\bf{y}}
\end{equation}
and variance
\begin{equation}\label{eq3.12}
{\bf{\sigma }}_{{\bf{\mathord{\buildrel{\lower3pt\hbox{$\scriptscriptstyle\frown$}} 
\over y'} }}}^{\bf{2}}{\bf{ = k(x',x') - k(x',x}}{{\bf{)}}^{\bf{T}}}{\bf{k(x,x}}{{\bf{)}}^{{\bf{ - 1}}}}{\bf{k(x',x)}}
\end{equation}
A comprehensive description and implementation of the convolution process can be found in \cite{alvarez2009sparse} and \cite{Alvarez2014} respectively. In this study, the convolution of two square exponential kernels are used since squared exponential kernel is a universal kernel \cite{RN173}, provided that data is stationary and the function to be modelled is a smooth one. Furthermore, squared exponential kernel has small quantity of hyperparameters to be established.
\subsection[Learning Hyperparameters]{Learning Hyperparameters}
There are two principal methods for learning the hyperparameters $\Theta $  namely: Bayesian model interference and marginal likelihood. Bayesian inference assumes that prior data of the unknown function to be mapped are known and a posterior distribution over the function is refined by incorporation of observations. The marginal likelihood method is based on the aspect that some hyperparameters are going to be more noticeable. Over this base, the posterior distribution of hyperparameters can be described with a unimodal narrow Gaussian distribution.

The learning of GPs hyperparameters  $\Theta $ is normally carried out with the maximization of marginal likelihood. The marginal likelihood can be expressed as:
\begin{equation}\label{eq3.13}
p\left( {{\bf{y}}\left| {{\bf{x}},{\bf{\Theta }}} \right.} \right) = \frac{1}{{{{\left( {2\pi } \right)}^{\frac{N}{2}}}{{\left| {\bf{K}} \right|}^{\frac{1}{2}}}}}{e^{ - \frac{1}{2}{{\bf{y}}^T}{{\bf{K}}^{ - 1}}{\bf{y}}}}
\end{equation}
where ${\bf{K}}$ is the covariance matrix,  $N$ is the number of input learning data points and ${\bf{y}}$ is a vector of learning output data of the form  $\left[ {{y_1};{y_2}; \cdots {y_N}} \right]$ . To reduce the calculation complexity, it is preferred to use the logarithmical marginal likelihood that is obtained by the application of logarithmic properties to \cref{eq3.14}.
\begin{equation}\label{eq3.14}
{\cal L}\left( {\bf{\Theta }} \right) =  - \frac{1}{2}\log \left( {\left| {\bf{K}} \right|} \right) - \frac{1}{2}{{\bf{y}}^T}{{\bf{K}}^{ - 1}}{\bf{y}} - \frac{N}{2}\log \left( {2\pi } \right)
\end{equation}
In order to find out a solution for the maximization of log-likelihood, there are multiple optimization methods that can be used such as particle swarm optimization, genetic algorithms, or gradient descent. For deterministic optimization methods, the computation of likelihood partial derivatives with respect to each hyperparameter is required. From \cite{williams2006gaussian} log-likelihood derivatives for each hyperparameter can be calculated by:
\begin{equation}\label{eq3.15}
\frac{{\partial {\cal L}\left( {\bf{\Theta }} \right)}}{{\partial {{\bf{\Theta }}_i}}} =  - \frac{1}{2}trace\left( {{{\bf{K}}^{ - 1}}\frac{{\partial {\bf{K}}}}{{\partial {{\bf{\Theta }}_i}}}} \right) + \frac{1}{2}{{\bf{y}}^T}{{\bf{K}}^{ - 1}}\frac{{\partial {\bf{K}}}}{{\partial {{\bf{\Theta }}_i}}}{{\bf{K}}^{ - 1}}
\end{equation}

\Cref{eq3.14} gives us the learning process computational complexity, for each cycle the inverse of the covariance matrix of ${\bf{K}}$ has to be calculated. This calculation carries a complexity ${\bf{O}}{\left( {{\bf{NM}}} \right)^3}$ . After learning, the complexity of predicting the value ${\bf{y}}(k + 1)$ is  ${\bf{O}}\left( {{\bf{NM}}} \right)$ and to predict the mean value  ${\bf{\sigma }}(k + 1)$ is  ${\bf{O}}{\left( {{\bf{NM}}} \right)^2}$ . The higher order term  ${\bf{O}}{\left( {{\bf{NM}}} \right)^{\bf{3}}}$ is the major disadvantage of using multi-output GPs. If the number of data increases the complexity of learning the hyperparameters increases in a cubic form. Methods such as genetic algorithms, differential equations, and particle swarm optimization can be applied to avoid the calculation of the marginal likelihood partial derivatives and thereby reduce the computational time.

\section[Experiment Setup and Results]{Experiment Setup and Results}
\subsection[Experiment setup]{Experiment setup}
\begin{figure}[!htb]
\centering
\includegraphics[scale=.4]{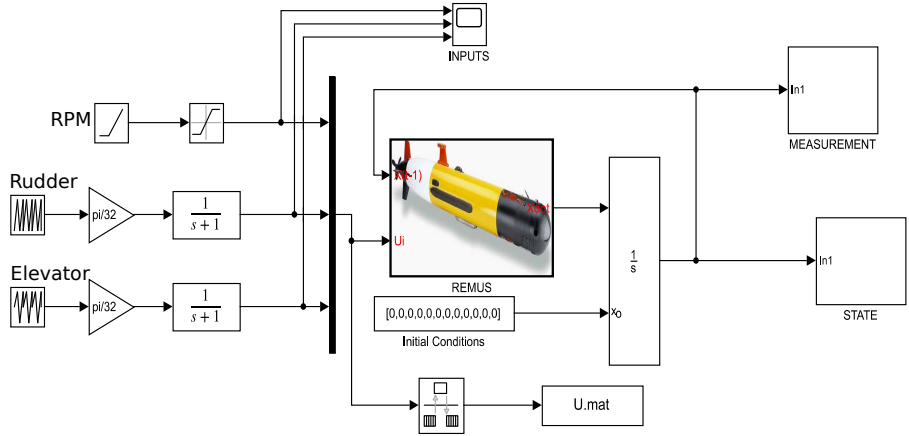}
\caption{REMUS 100 AUV Simulink model.}
\label{fig:figure-2}
\end{figure}

\begin{figure}[!htb]
\centering
\includegraphics[scale=.4]{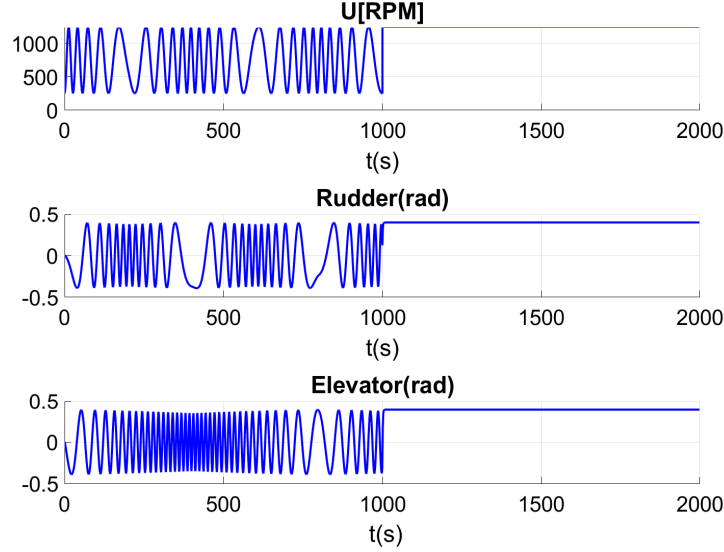}
\caption{Remus 100 input signals example \textit{chirp + step} for all inputs.}
\label{fig:figure-3}
\end{figure}

The implementation of a mathematical model of an underactuated REMUS 100 AUV was used to generate the required identification data. The coefficients of \cite{RN178} were used and adapted for simulation on Simulink with the addition of the thruster model from \cite{RN163} as the original mathematical model produced by \cite{RN178} has a constant thrust force. The resultant model was tested to mimic the original results obtained by \cite{RN178} at a speed of 1.5 m/s. The AUV details can be found in \cref{table3.1}. As shown in \cref{fig:figure-2} a simulation setup was developed in MATLAB/Simulink to emulate the AUV behaviour. \Cref{fig:figure-3} shows an example of input signals given for the rudder angle, thruster RPMs and elevator angle respectively. A total of 8 sets of data were produced by combining and initial chirp signal and after the first 1000 seconds the command signal change to a step function or ramp function. Simulation was carried out for 2000 seconds. A list of simulation done can be seen in \cref{tabblesimulations}. The objective of not using a standard test such as zigzag test or turning circle test is to test the ability of GPs under more drastic conditions. A sample data point was captured for each 1.5 seconds over the input and outputs. A total of 8000 points were captured over the six motion outputs and 4000 point over the three input signals (Propeller RPM, rudder angle and elevator angle). The data set was divided into two sets of points, the first set of points is used for the model learning, this data of training is equivalent to the section of chirp input signal, and the second set of points is used for learning validation. The validation data is purposely chosen to be beyond the range of training data and very different to the training data to test the ability of the method to predict beyond the training range.

\begin{table}[]

\centering
\caption{REMUS 100 general characteristics.}
{\begin{tabular}{l|l}
\multicolumn{1}{c|}{\textbf{Parameter}} & \multicolumn{1}{c}{\textbf{Value}} \\ \hline
Weight                                  & 299(N)                             \\
Buoyancy                                & 306(N)                             \\
Vehicle total length                    & 1.33(m)                            \\
Diameter                                & 0.191(m)                              \\
Max. Depth                              & 100(m)                            
\end{tabular}}
\label{table3.1}
\end{table}

\begin{table}[htbp]
\centering
\centering\caption{Simulation description} 
\centering
\begin{tabular}{c|c}
\multicolumn{1}{c|}{\textbf{Experiment Number}} & \multicolumn{1}{c}{\textbf{Experiment Configuration}} \\

    \midrule
    1     & Chirp+Ramp in Propeller \\
    2     & Chirp+Ramp in Rudder \\
    3     & Chirp+Ramp in Stern \\
    4     & Chirp+Ramp in All surfaces \\
    5     & Chirp+Step in Propeller \\
    6     & Chirp+Step in Rudder \\
    7     & Chirp+Step in Stern \\
    8     & Chirp+Step in All surfaces \\
    \end{tabular}%
    
\label{tabblesimulations}%
\end{table}%

\subsection[Training and validation]{Training and validation}
A script was written to implement the NARX structure with multi-output GPs and the implementation of multi-out GPs by \cite{Alvarez2014} was employed. The multi-output GPs regressors were defined as:
\begin{equation}\label{eq3.16}
\left[ {\begin{array}{*{20}{c}}
{{{\bf{u}}_k}}\\
{{{\bf{v}}_k}}\\
{{{\bf{w}}_k}}\\
{{{\bf{p}}_k}}\\
{{{\bf{q}}_k}}\\
{{{\bf{r}}_k}}
\end{array}} \right] = f\left( {{\bf{y}},{{\bf{c}}_{k - 1:3}}} \right)
\end{equation}

Where ${\bf{y}}$ is the vector of regressors  ${\left[ {{{\bf{u}}_{k - 1:3}},{{\bf{v}}_{k - 1:3}},{{\bf{w}}_{k - 1:3}},{{\bf{p}}_{k - 1:3}},{{\bf{q}}_{k - 1:3}},{{\bf{r}}_{k - 1:3}}} \right]^T}$ , the function  $f$ is a relation between the vector of regressors from the correspondent vehicle speeds  $\left( {{\bf{u,v,w,p,q,r}}} \right)$ or the full vehicle state ${\bf{y}}$  , and the vector  ${\bf{c}}$that content the regressors of the commanded signals $\left( {{u_{rpm}},{u_{elevator}},{u_{rudder}}} \right)$  to the respective output of the system. The input signals were normalized between -1 and 1 to give all the inputs and outputs the same weight in the learning process. 

For the training, a minimum search with the gradient descent method, in particular the interior-point algorithm, was used for the minimization of the negative logarithmical likelihood. 

\begin{figure}[!htb]
\centering
\includegraphics[scale=.5]{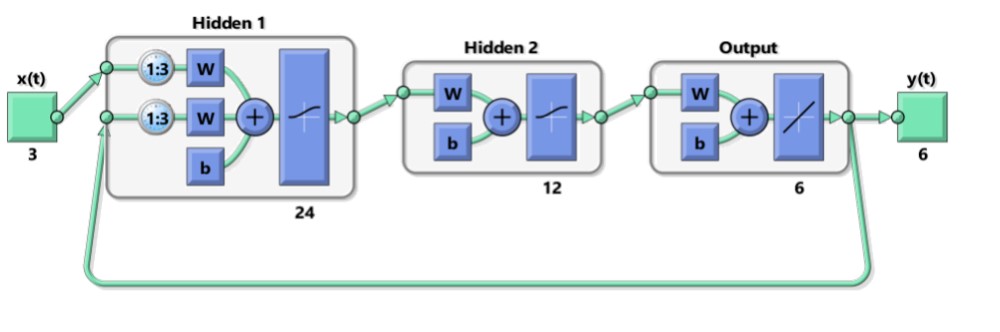}
\caption{\textit{RNN 1} configuration for AUV identification.}
\label{fig:figure-4}
\end{figure}

\begin{figure}[!htb]
\centering
\includegraphics[scale=.5]{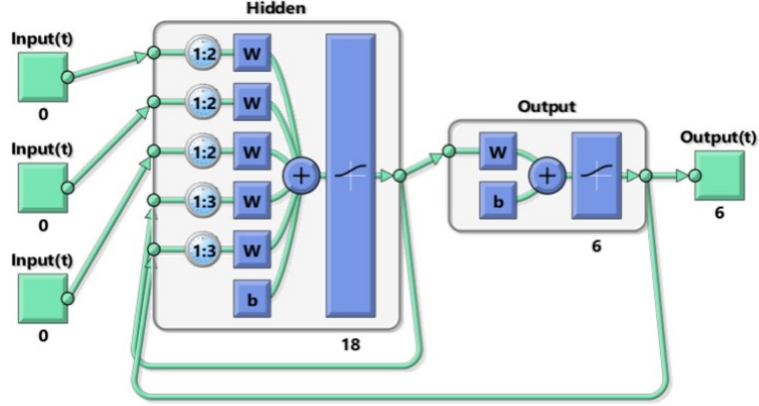}
\caption{\textit{RNN 2} configuration for AUV identification.}
\label{fig:figure-4b}
\end{figure}

\begin{figure}
\centering
\subfloat{\includegraphics[width=75mm]{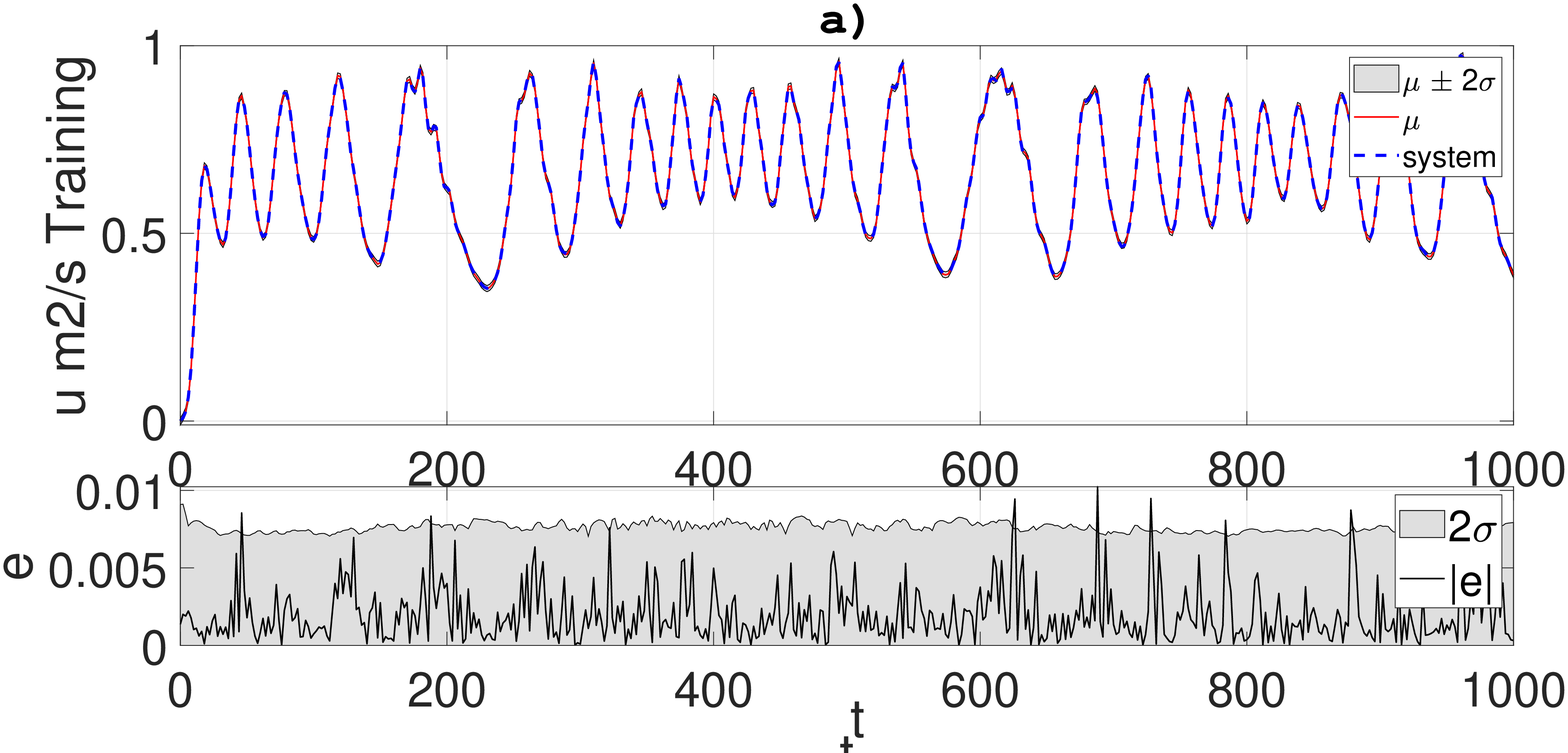}}
  \subfloat{\includegraphics[width=75mm]{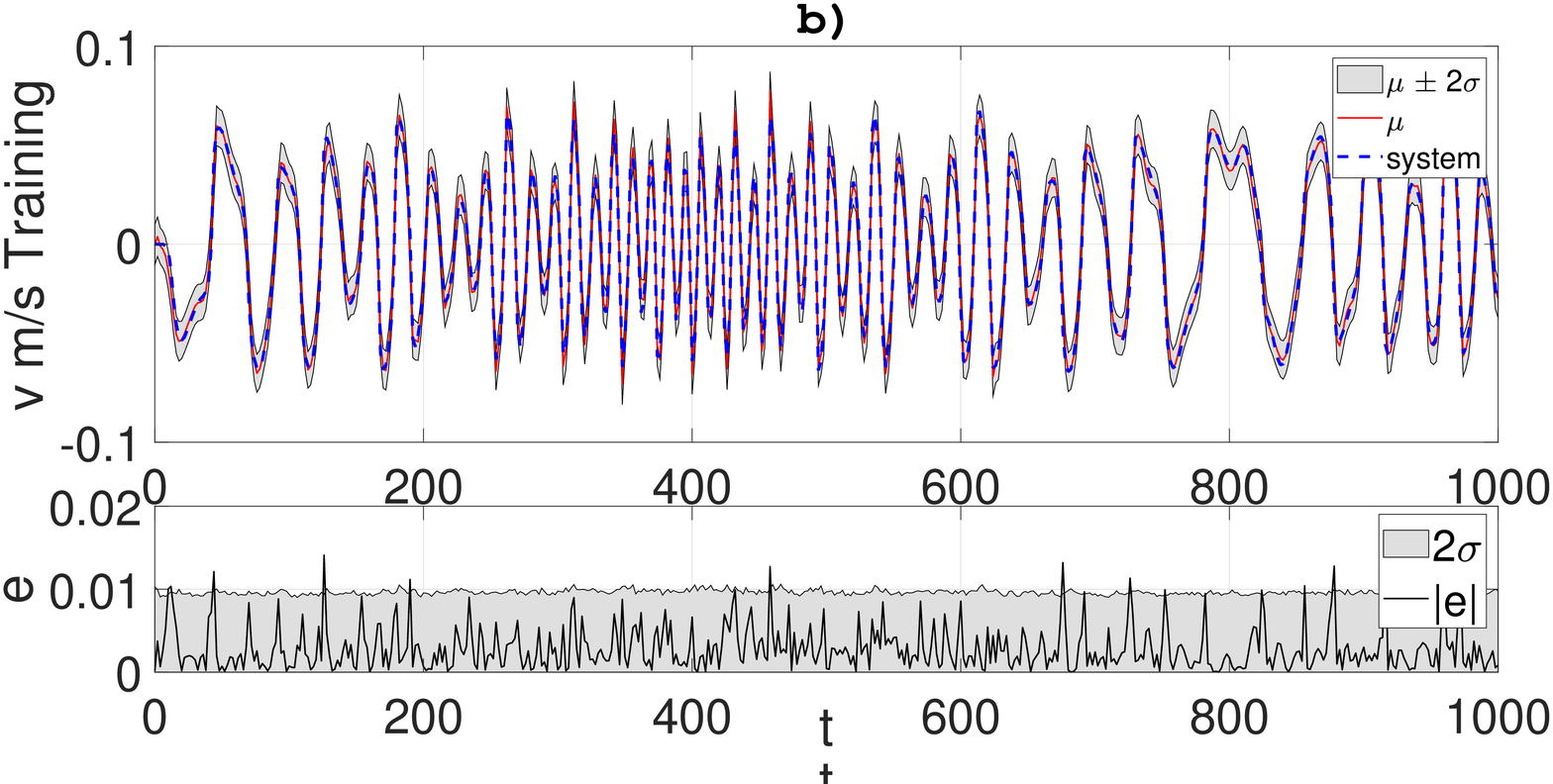}}
  \\
  \subfloat{\includegraphics[width=75mm]{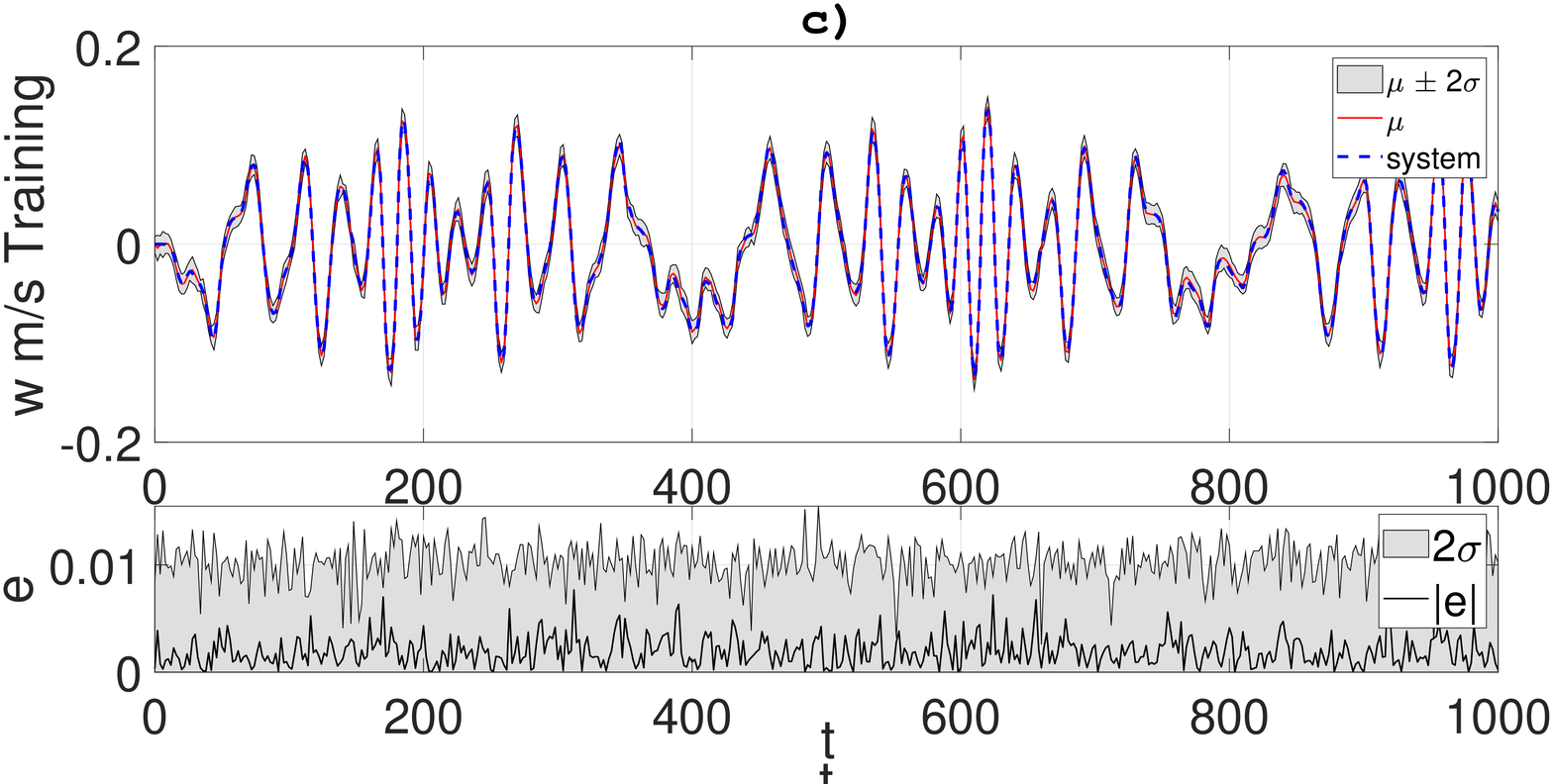}}
  \subfloat{\includegraphics[width=75mm]{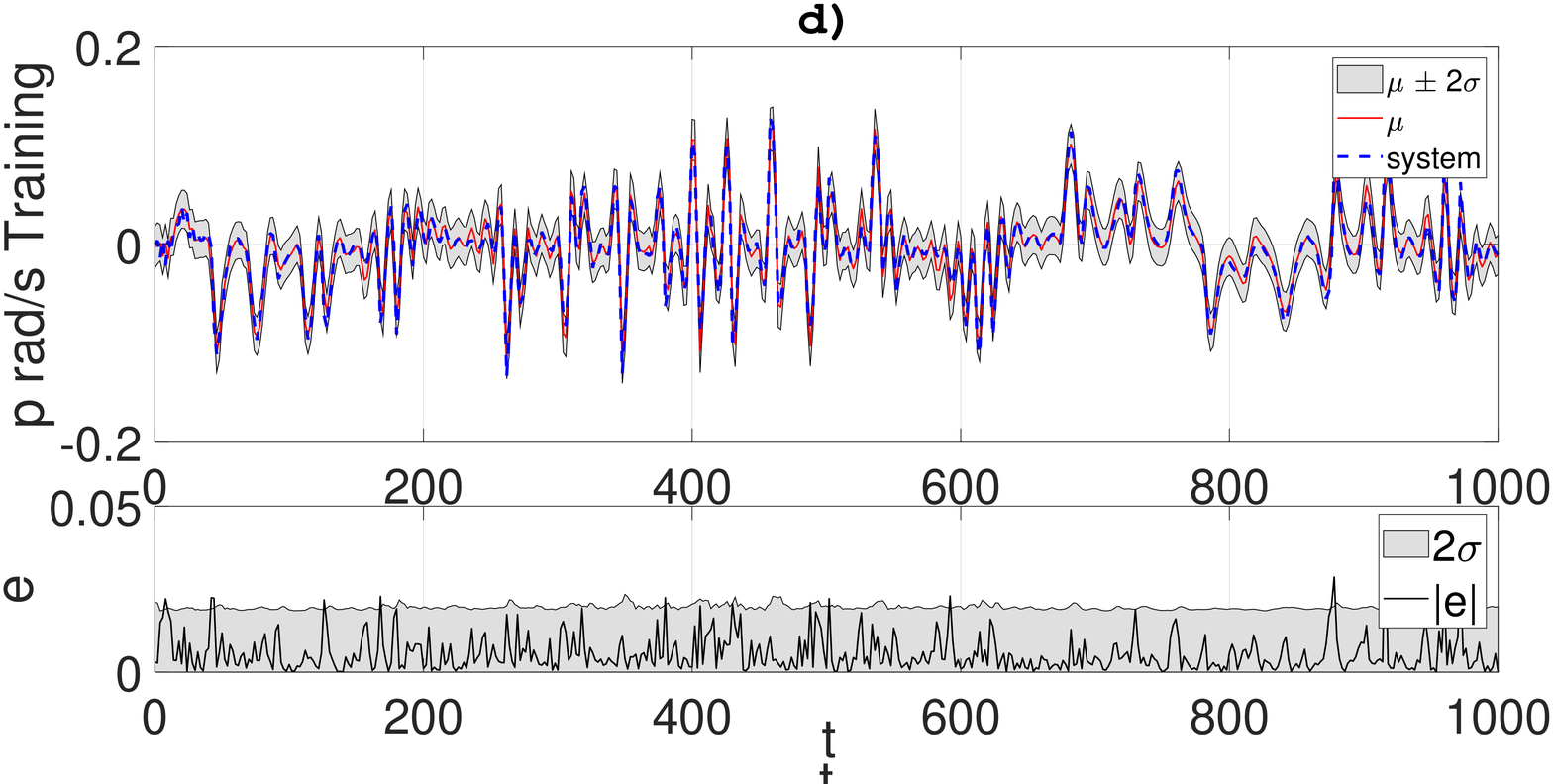}}
  \\
  \subfloat{\includegraphics[width=75mm]{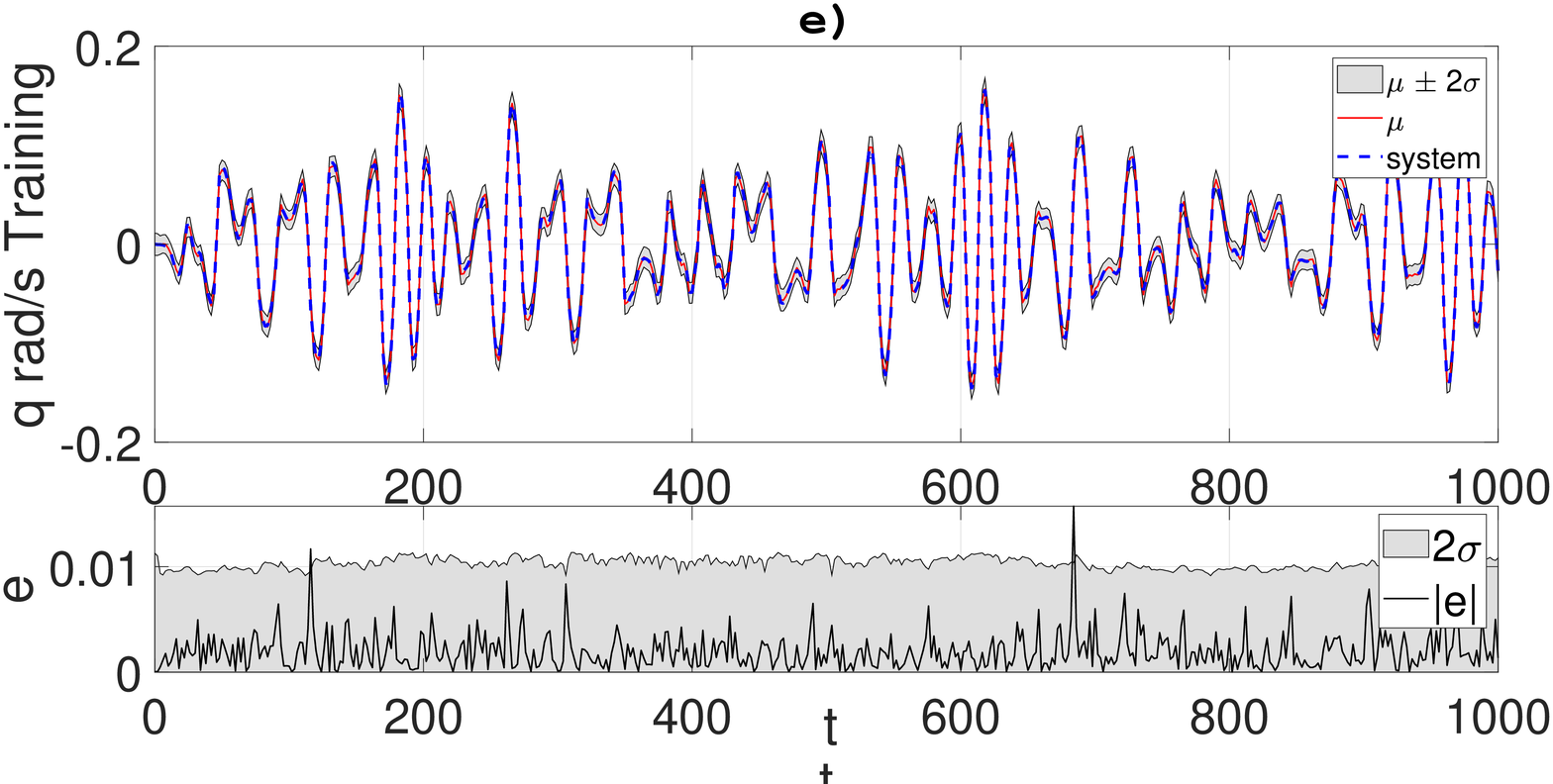}}
  \subfloat{\includegraphics[width=75mm]{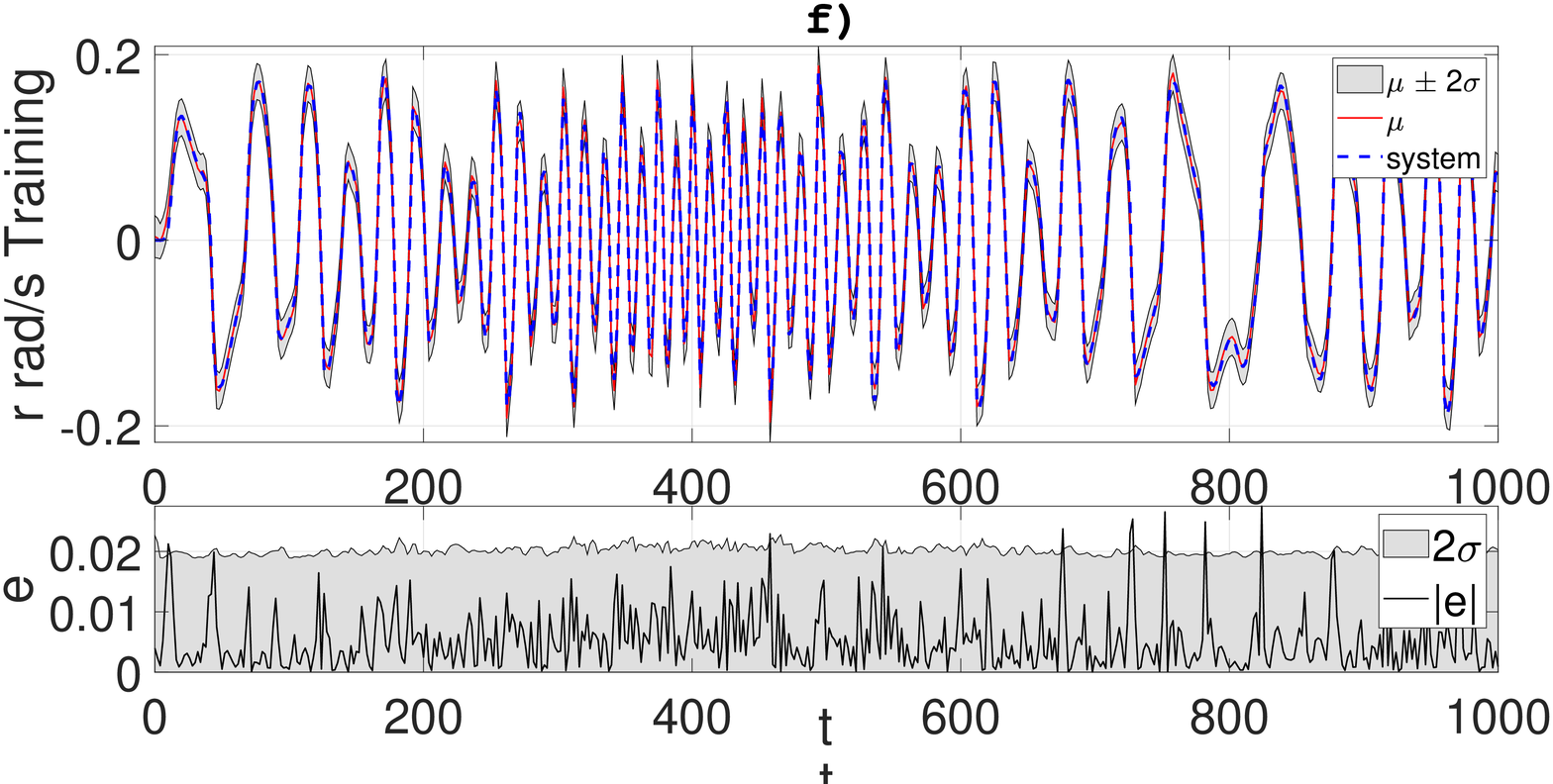}}
\caption{Multi-output GPs training, a) surge velocity, b) sway velocity, c) heave velocity, d) roll angular velocity, e) pitch angular velocity and f) yaw angular velocity.} \label{fig:figure-5}
\end{figure}

Two neural network nonlinear system identification model for comparison were also prepared. The NN systems are recurrent neural network (RNN) as shown in \cref{fig:figure-4,fig:figure-4b}. The first MIMO RNN (\textit{RNN 1},\Cref{fig:figure-4}) was setup with three terms of delays for the output to be feedback to the network and three terms delay of the inputs. The \textit{RNN 1} that was selected as relatively optimal for the task at hand used two hidden layers with logarithmic sigmoid functions for the hidden layer neurons and was trained with Levenberg-Marquardt backpropagation, this configuration and regressors was selected as provide the best results for our system. \textit{RNN 2}(\Cref{fig:figure-4b}) is a fully connected RNN with a single hidden layer. \textit{RNN 2} use the same logarithmic sigmoid functions for the hidden layer neurons and was trained with Levenberg-Marquardt backpropagation.  A third step of simulation was carry out with the combination the full length of the data and feeding back after each step the output from the models. The neural network system was trained, validated, and tested with the same data used for the multi-output GPs. The complete implementation code can be found at the GitHub Repository \footnote{https://github.com/ArizaWilmerUTAS/System-indetification-of-underwater-vehicles-with-Multi-Output-Gaussian-Processes}. \Cref{fig:figure-5} presents the results of GPs training compared to the AUV simulator signals, and the error plots between the predicted and real systems. In all the figures, a  $2\sigma $ variance is plotted. The variance values of the training data are in the expected value and encompass the error results.

\begin{figure}
\centering
  \subfloat{\includegraphics[width=75mm]{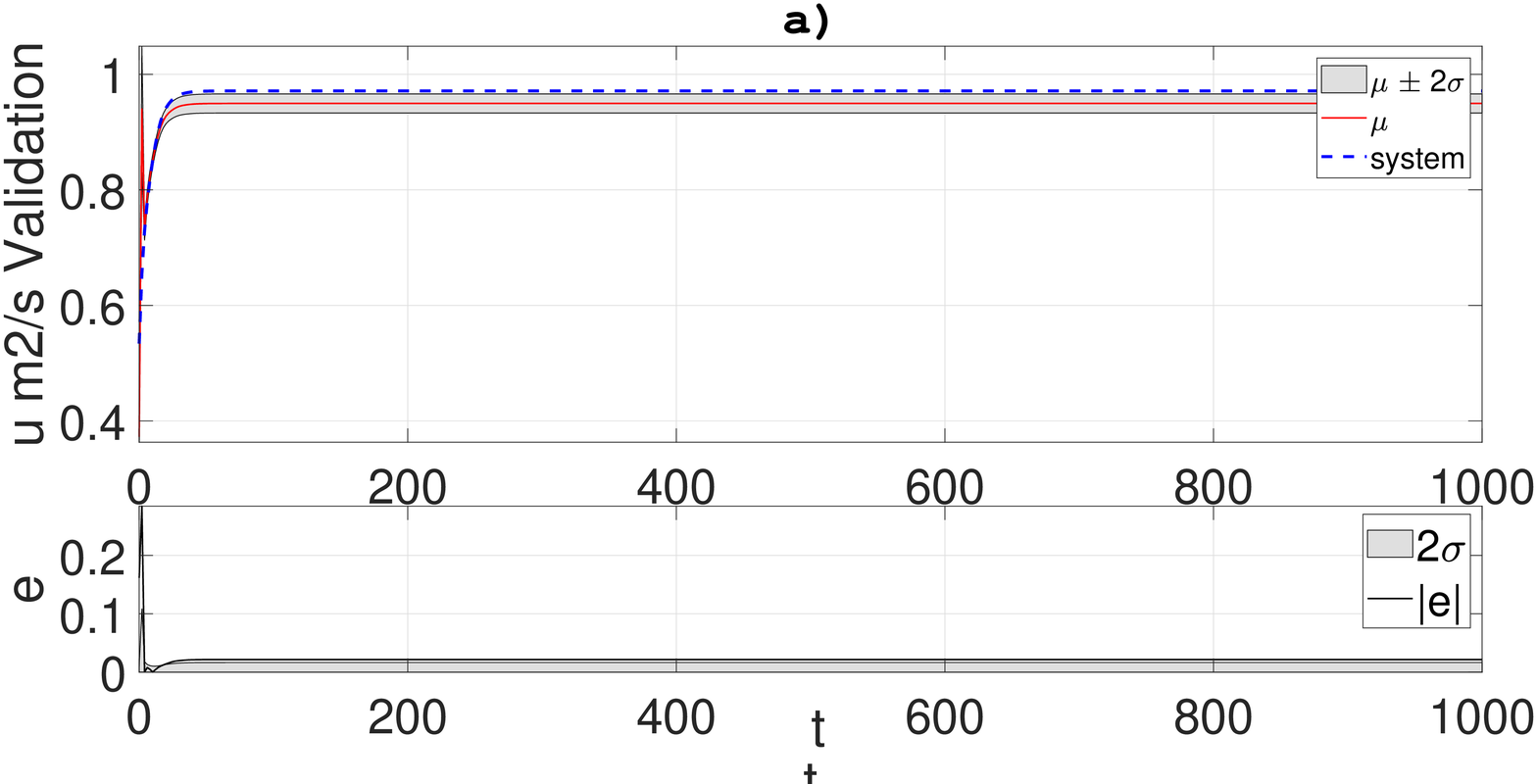}}
  \subfloat{\includegraphics[width=75mm]{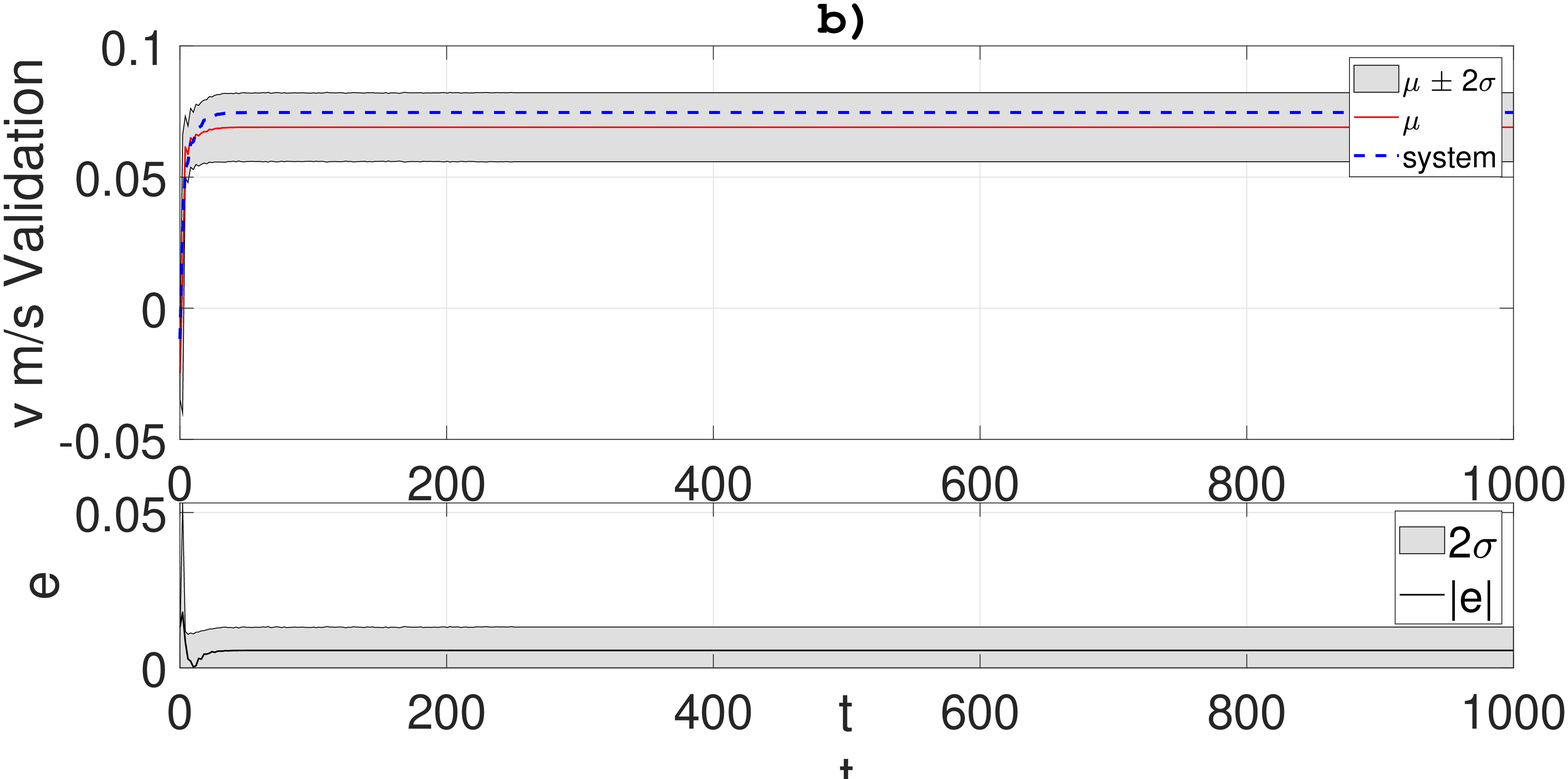}}
  \\
  \subfloat{\includegraphics[width=75mm]{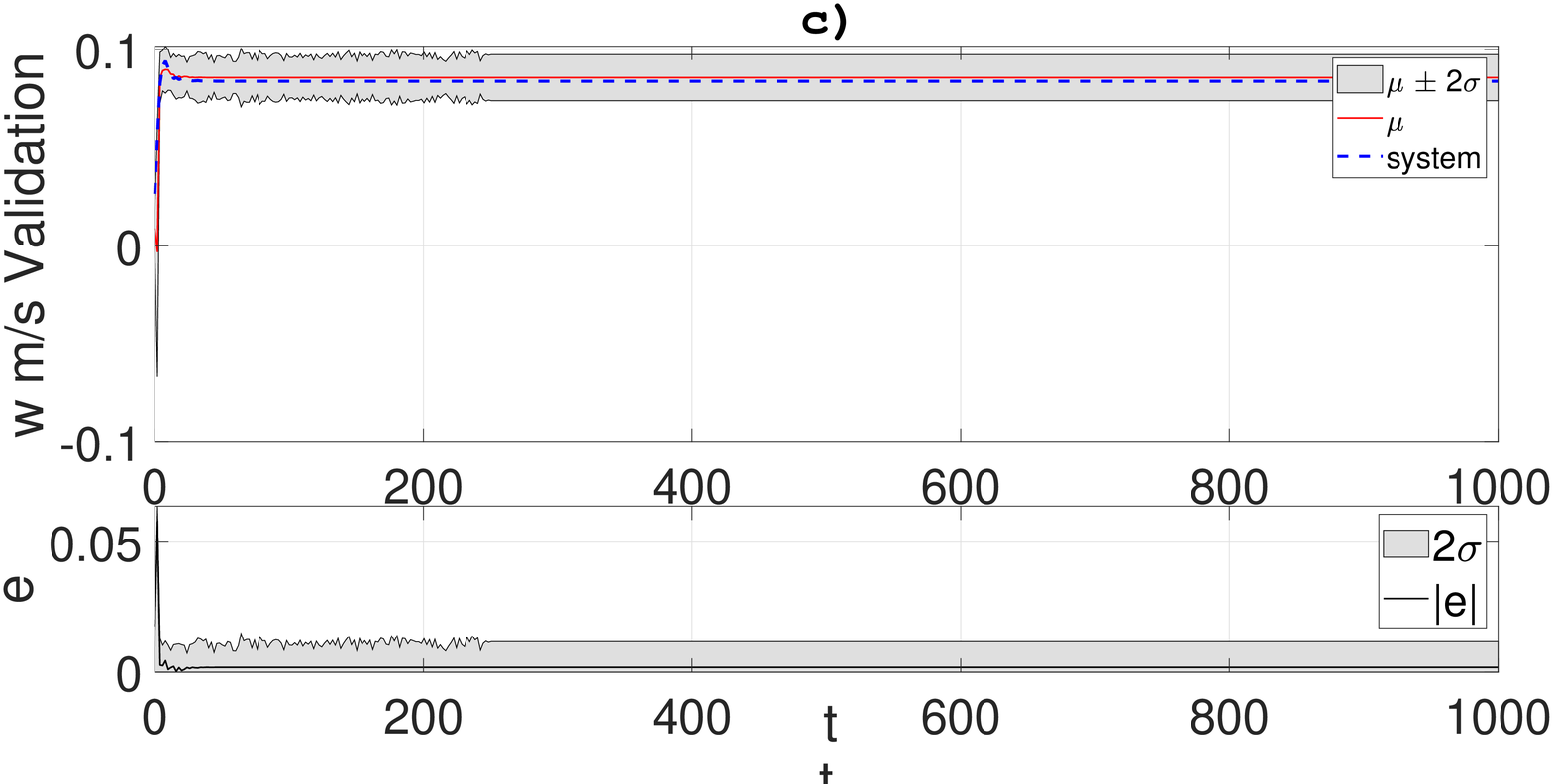}}
  \subfloat{\includegraphics[width=75mm]{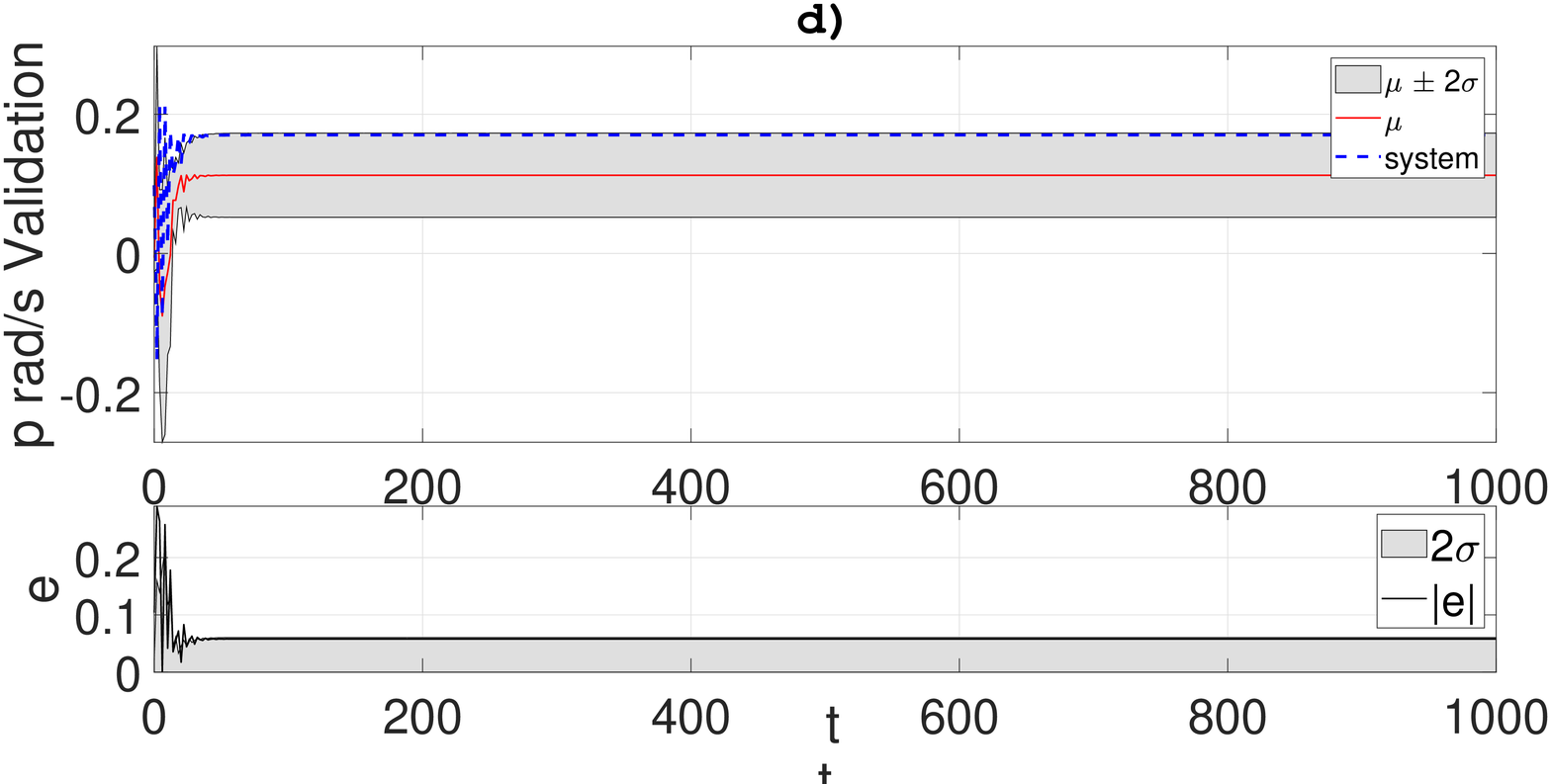}}
  \\
  \subfloat{\includegraphics[width=75mm]{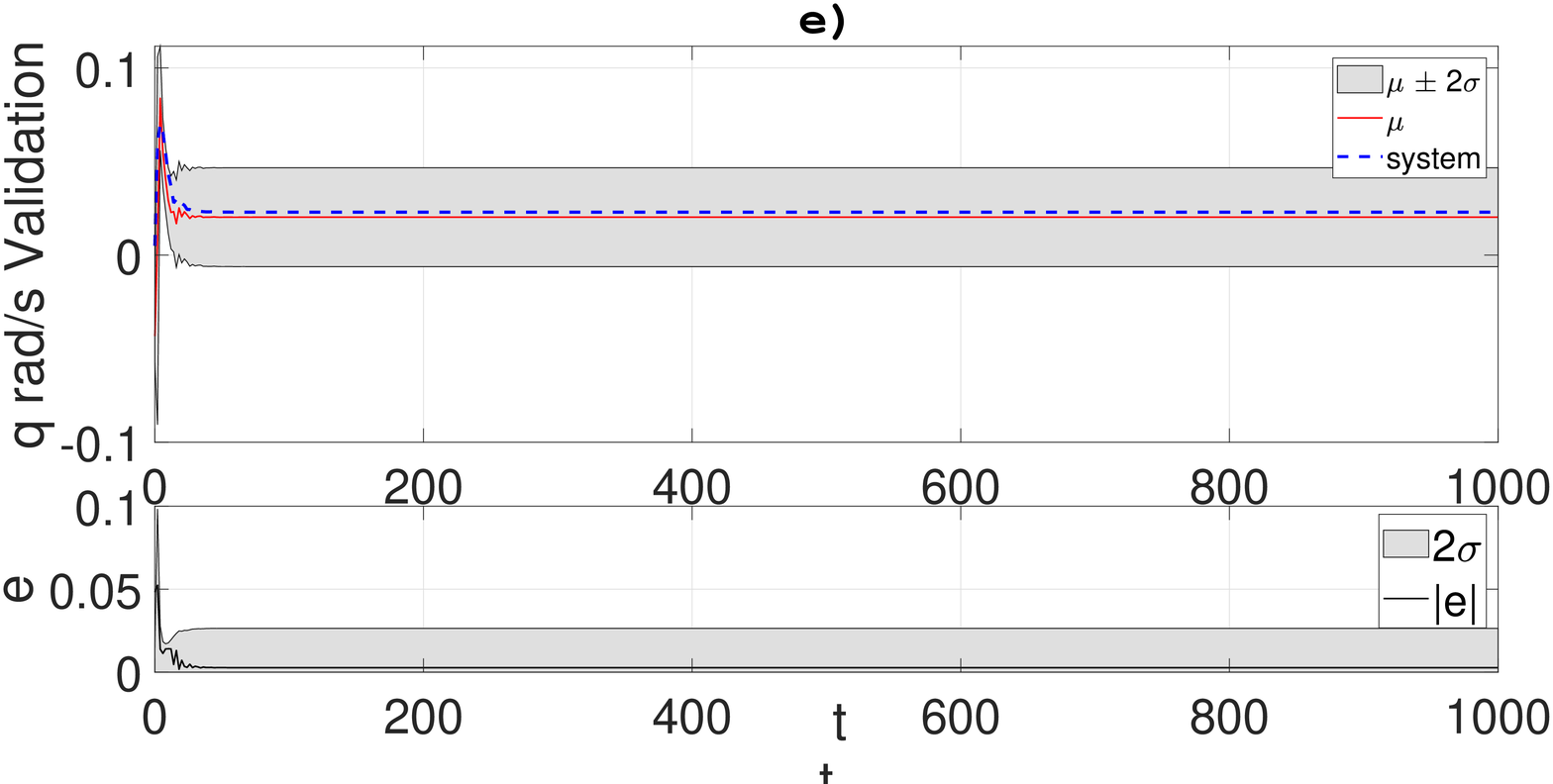}}
  \subfloat{\includegraphics[width=75mm]{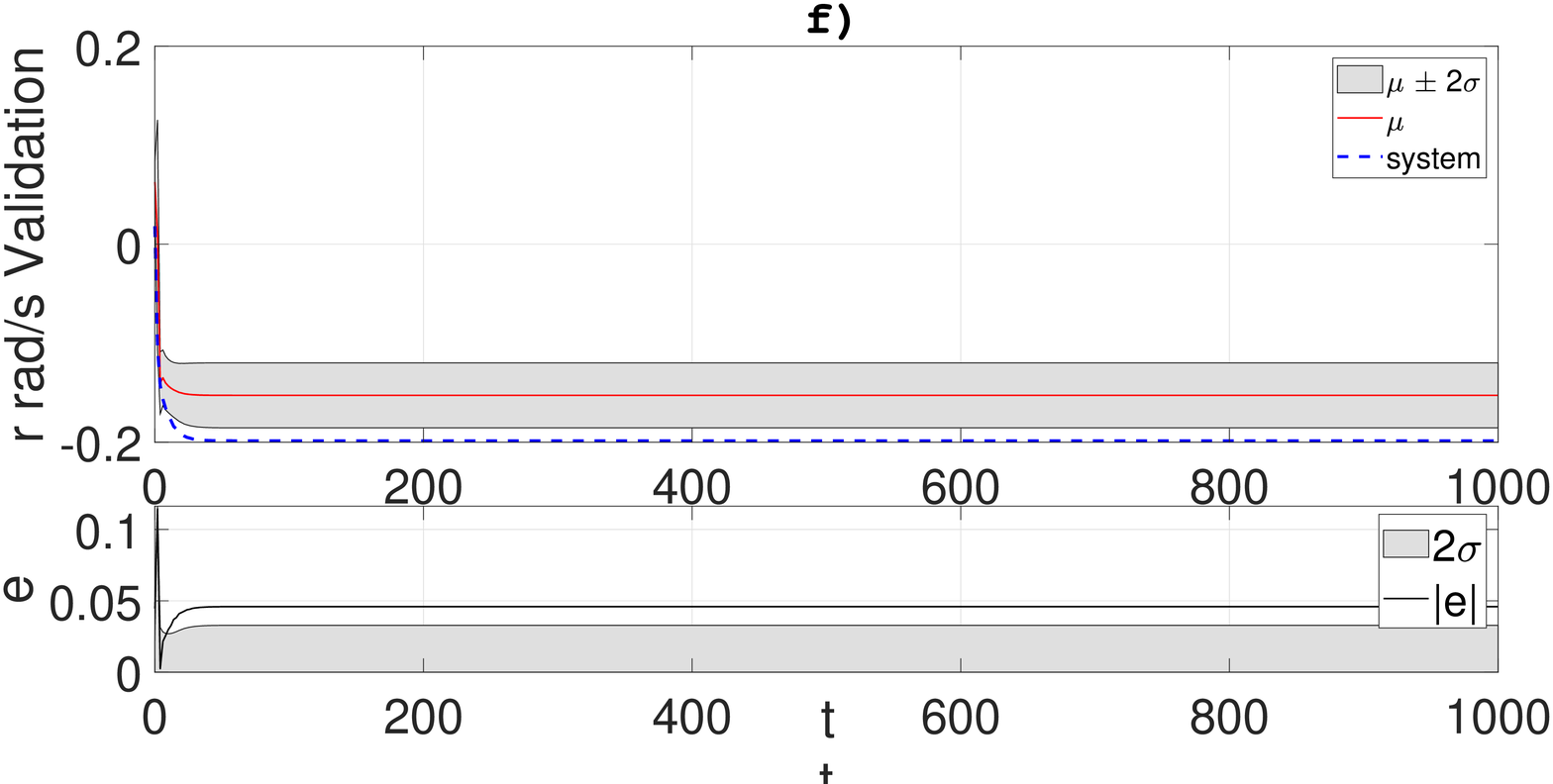}}
\caption{Multi-output GPs validation with unknown data, a) surge velocity, b) sway velocity, c) heave velocity, d) roll angular velocity, e) pitch angular velocity and f) yaw angular velocity.} \label{fig:figure-6}
\end{figure}

The validation data consisted of the second part of the captured data in the vector form with the commanded inputs and the real output from the training data with the respective system delays. The segments of results from the validation with the second set of data are depicted in \cref{fig:figure-6}. The low validation errors show a good system prediction for the sway speed and yaw speed. 

\subsection[Simulation]{Simulation}
\begin{figure}
\centering
  \subfloat{\includegraphics[width=75mm]{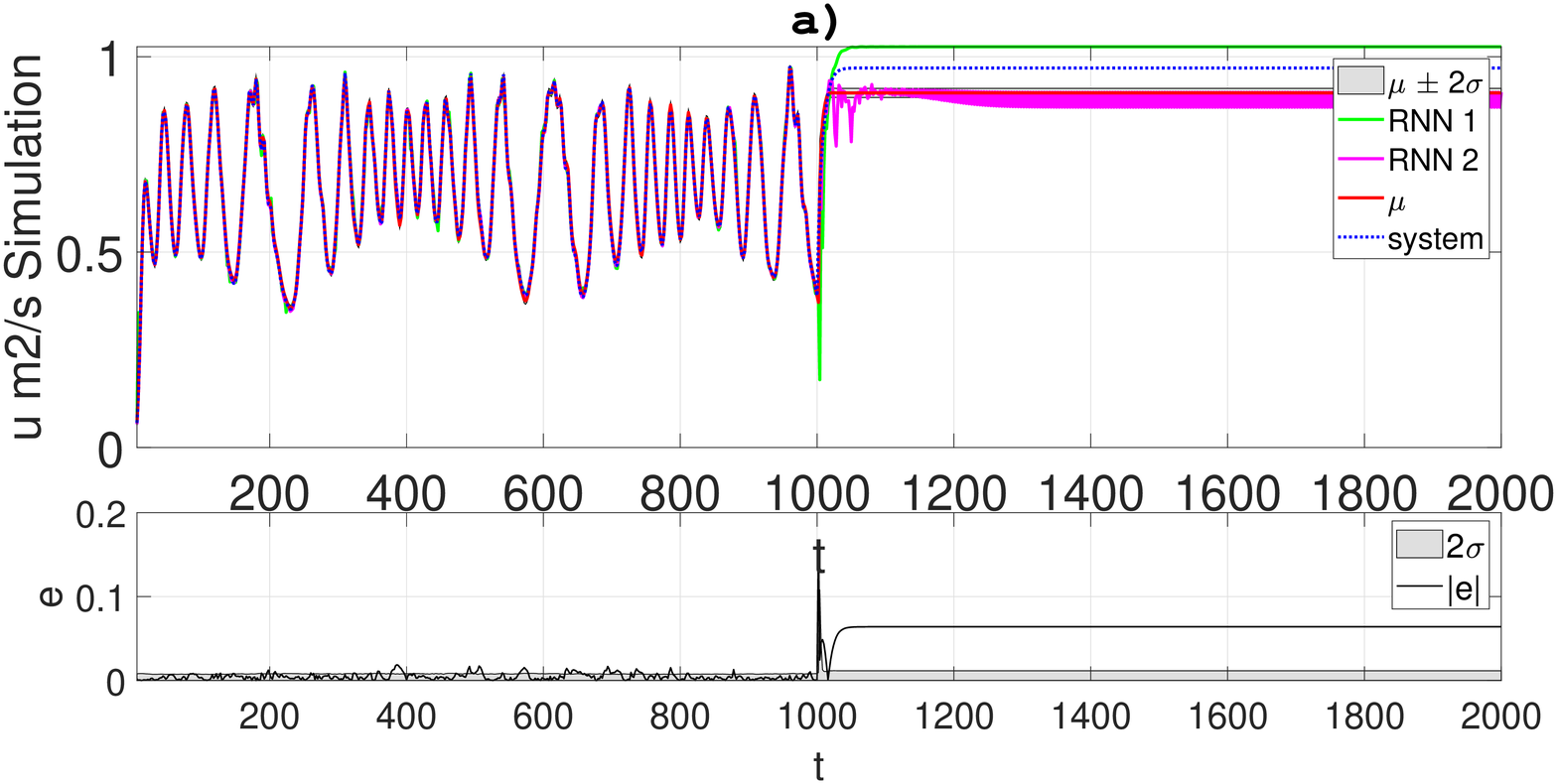}}
  \subfloat{\includegraphics[width=75mm]{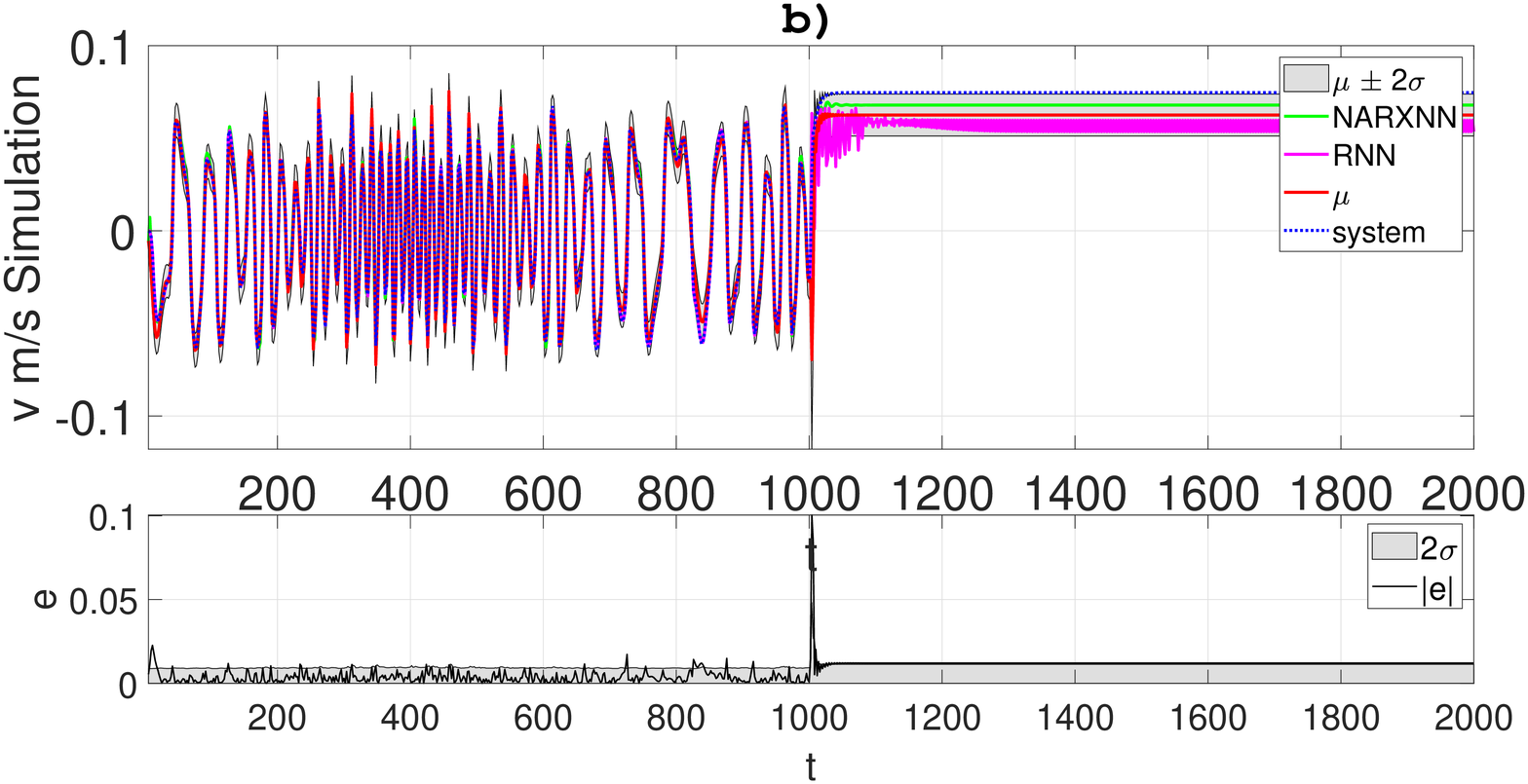}}
  \\
  \subfloat{\includegraphics[width=75mm]{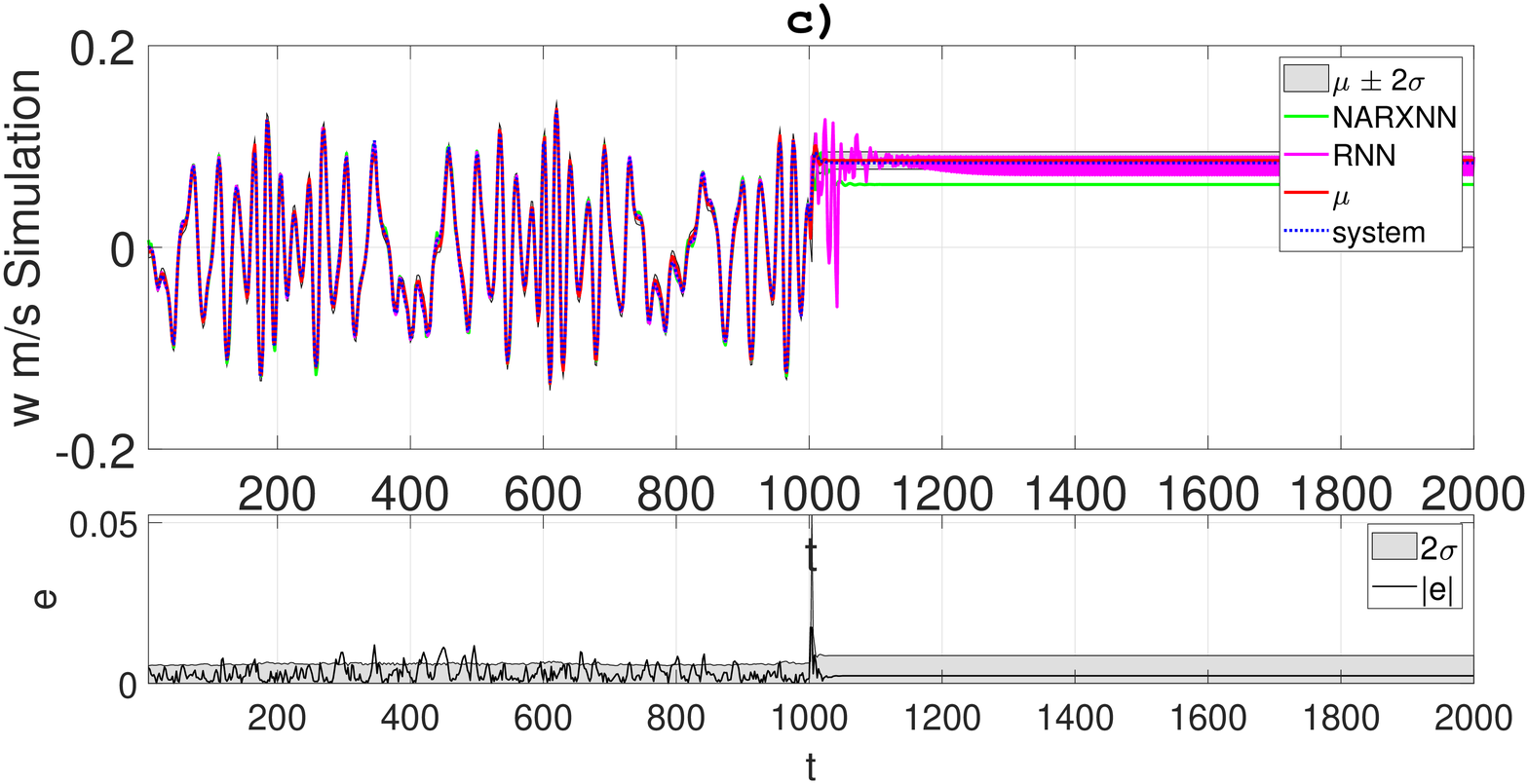}}
  \subfloat{\includegraphics[width=75mm]{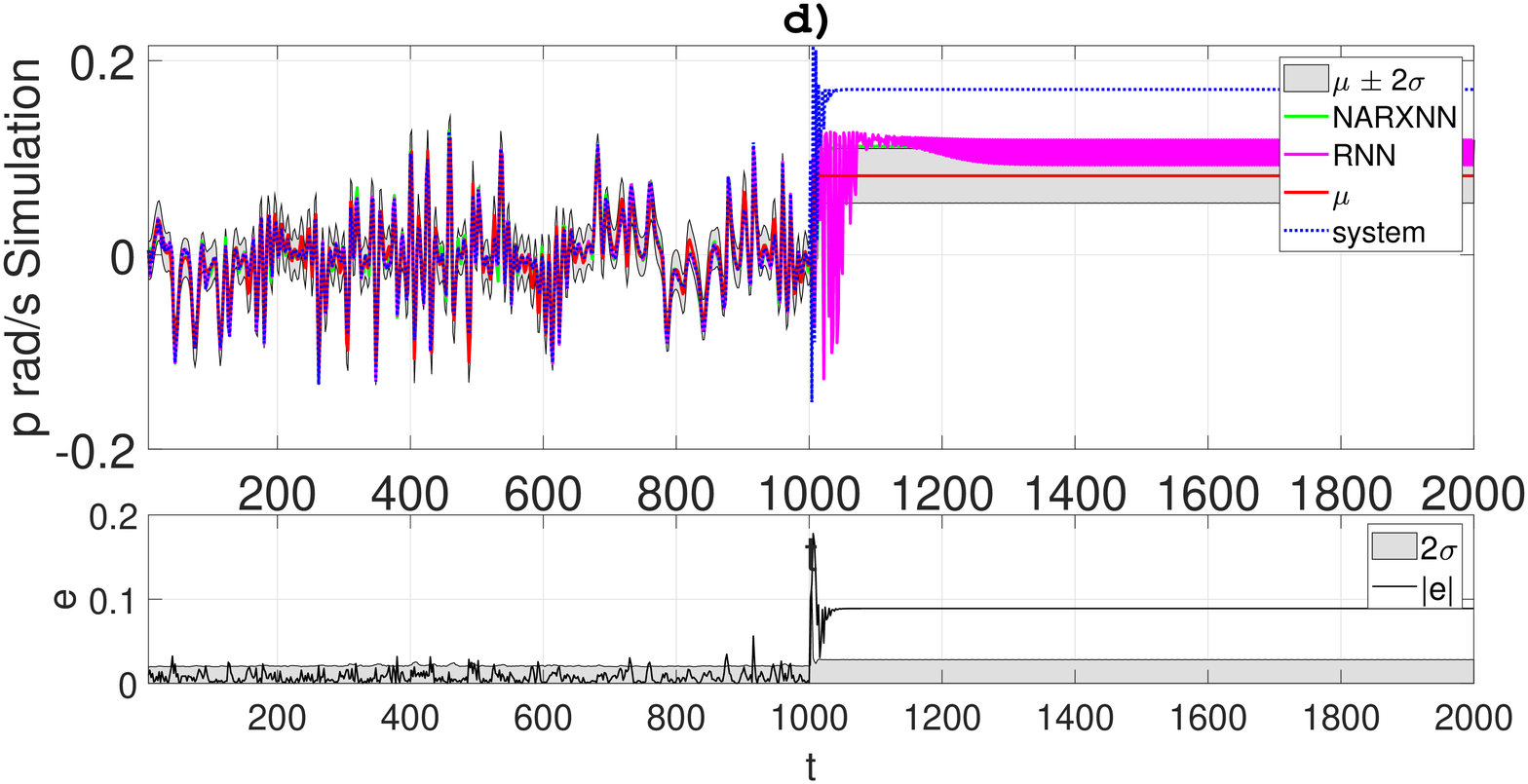}}
  \\
  \subfloat{\includegraphics[width=75mm]{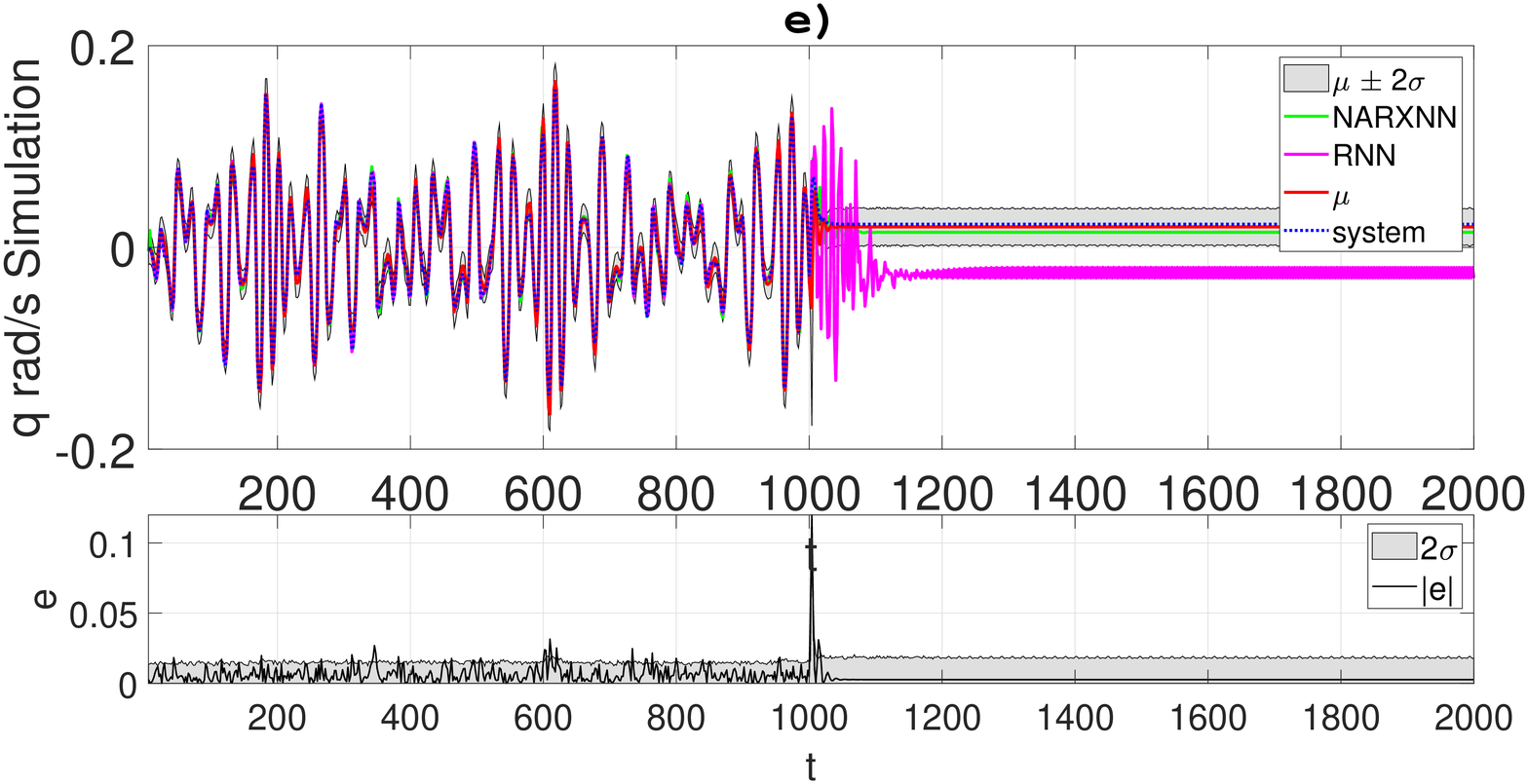}}
  \subfloat{\includegraphics[width=75mm]{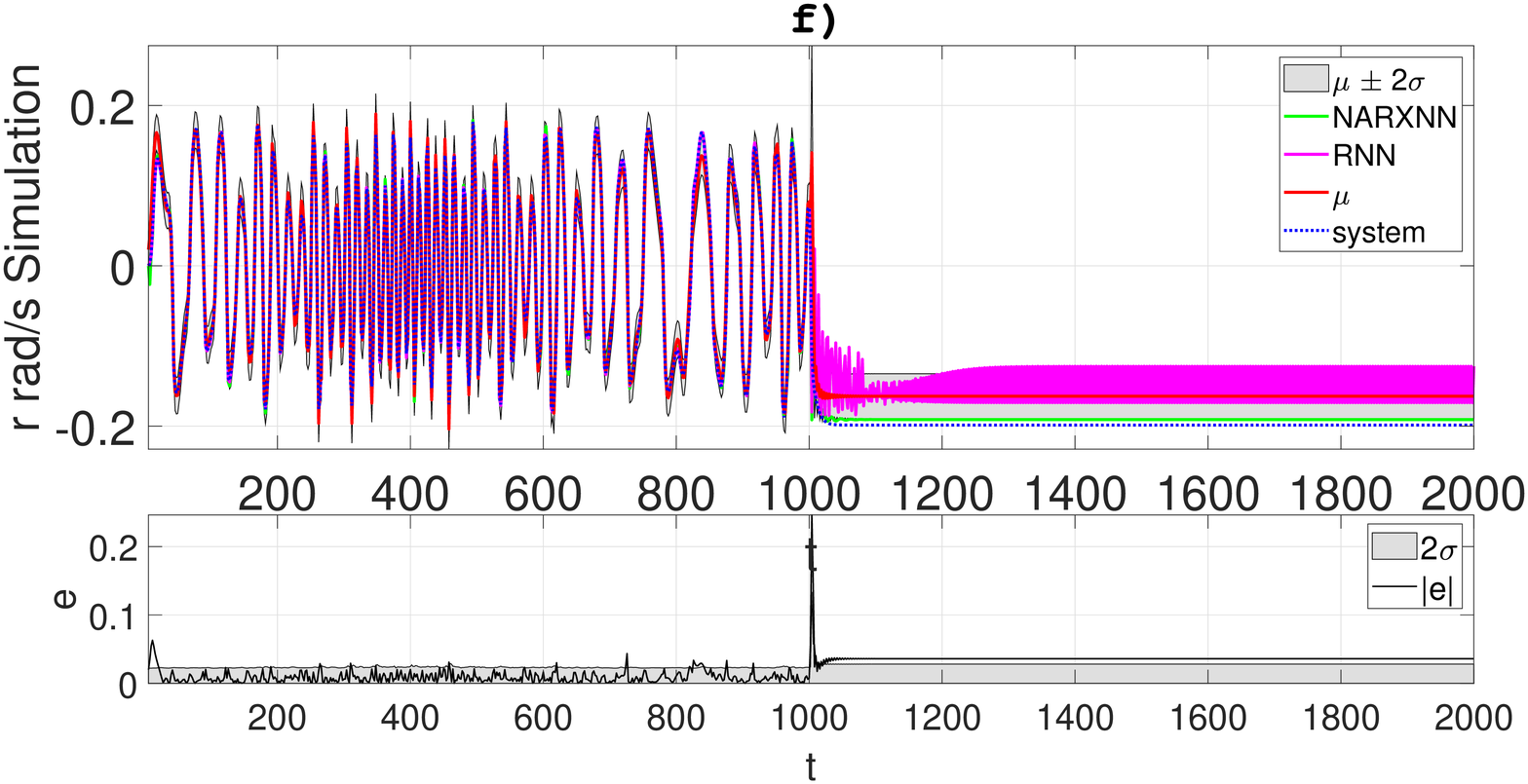}}

\caption{Multi-output GPs simulation compared with \textit{RNN 1}, \textit{RNN 2}, and real system with \textit{chirp+ramp} in all inputs a) surge velocity, b) sway velocity, c) heave velocity, d) roll angular velocity, e) pitch angular velocity and f) yaw angular velocity.} 
\label{fig:figure-7}
\end{figure}

With the objective to test the ability of the learning system, a simulation stage was implemented for the learned RNN and the multi-output Gaussian processes over the total length of the simulated data. Navigation applications as EKF, UKF, and control as model predictive control requires predicting the behaviour of the plant in a number steps ahead of the actual state to predict the correct position or control signals. In the case of the multi-output GPs the simulation is done by feeding back to the simulation the past inputs  ${y_i}(k - n)$ , the initial position and control signals of rudder, elevator and forward speed where used, the naive simulation \cite{RN81} covers training and validation data acquired from the original simulation in a close loop setup. Naive simulation provides an approximation where variances are not exactly the same, but provide a general guidance on uncertainty and this is enough for our study. If the variance is to be employed, such as in a control systems or a navigation problem, the uncertainty propagation can be included with the use a simulation based on Monte Carlo numerical approximation \cite{RN81}. \Cref{fig:figure-7} shows the results from the simulation of \textit{RNN 1}, \textit{RNN 2} and the multi-output GPs compare to the original system. \textit{RNN 1} compare to \textit{RNN 2} show better performance with ramp input signals, and  \textit{RNN 2} show better perfomance to simulate step functions in our simulations. However multi-output GPs can identify the system correctly and predict the behaviour of the system with \textit{chirp+ramp} and \textit{chirp+step} functions . The better capability of GPs to predict outside the training horizon from a number of difference variation from the initial training data is confirmed by the results of \cref{tab:rmse,tab:press,tab:mae}. The mean value of the output root-mean-square error (RMSE) ,the predicted residual error sum of squares (PRESS) and mean absolute error (MAE) for GPs are smaller than the values of \textit{RNN 1} and \textit{RNN 2}. 

\begin{figure}
	\centering
	\includegraphics[width=100mm]{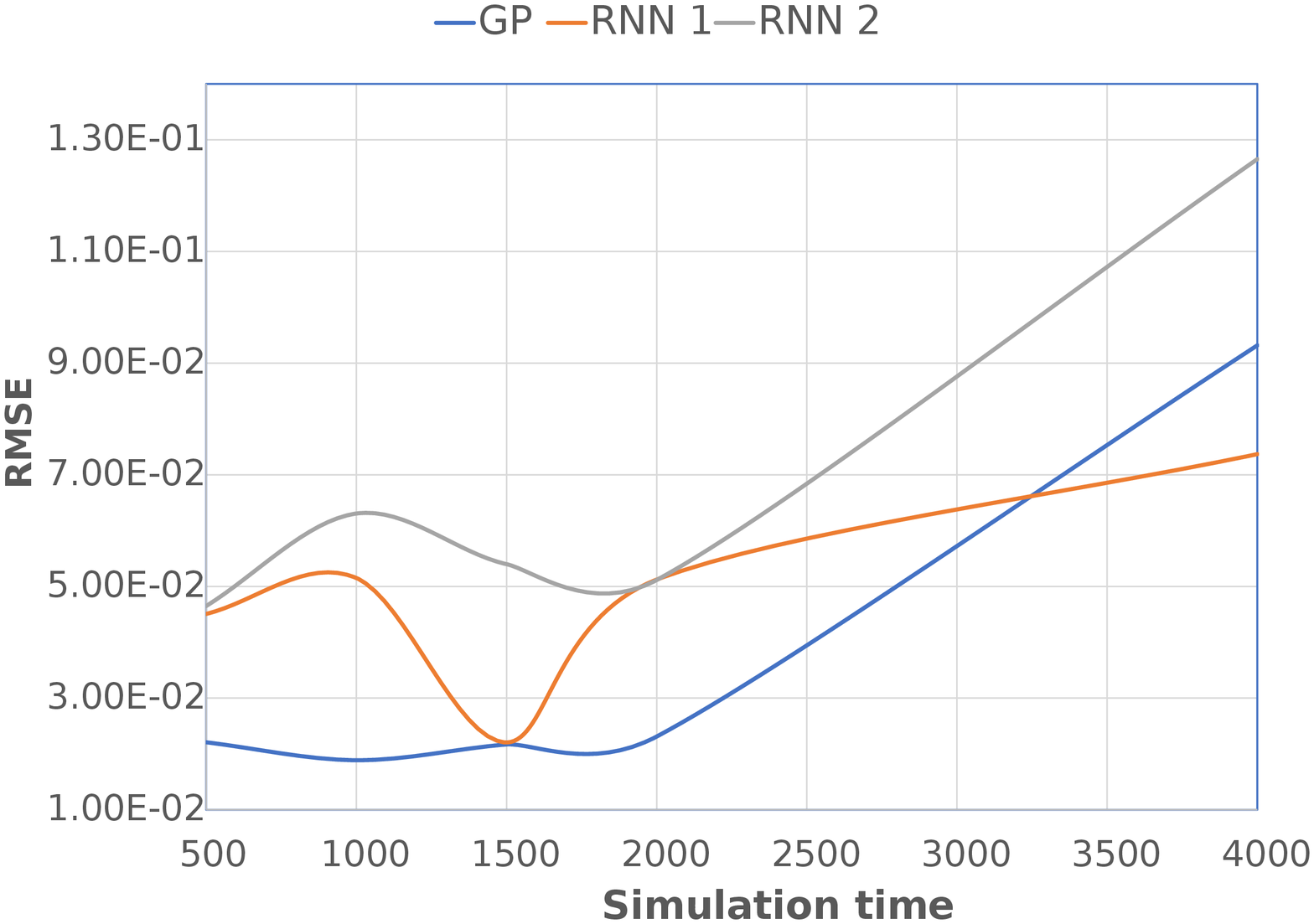}
	\caption{Training size sensitivity analysis RMSE results}
	\label{fig:rmse}
\end{figure}

\begin{figure}
	\centering
	\includegraphics[width=100mm]{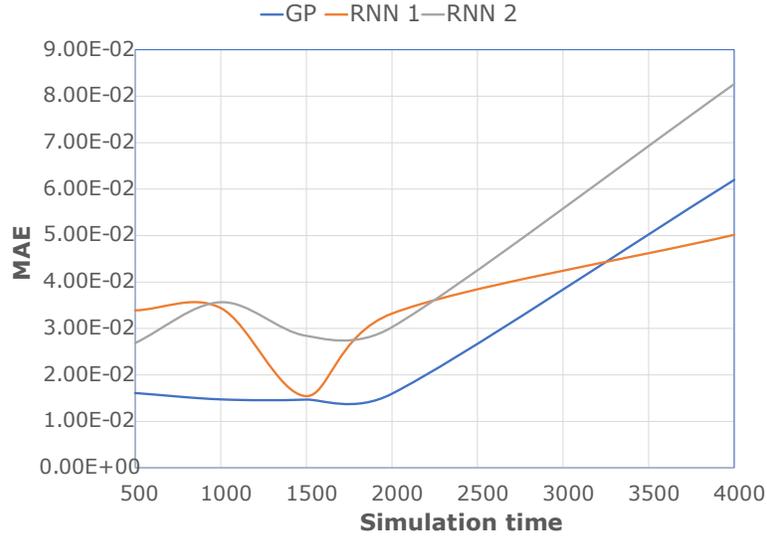}
	\caption{Training size sensitivity analysis MAE results}
	\label{fig:mae}
\end{figure}

A secondary set of simulations were also carried out to research the sensibility of multi-output GPs in comparison to both RNN in respect to the increase in training data. A series of simulations at 500, 1000, 1500,2000 and 4000 seconds were carried out with commands signals composed of a chirp signal for half of the time and a ramp signal for the other half of the simulation. The first half of each data set was employed for training and the test simulation of the learned model was done over the complete extension of data. The RMSE ,PRESS and MAE were also measured over all simulation results. The result of the sensitivity analysis can be seen in
\cref{fig:rmse,fig:mae}. All the measurement of the sensitivity analysis show the same trend for each measured variable, the simulations with 1500 seconds of capture data for \textit{RNN1} and GPs show similar average results and \textit{RNN 1} over 4000 seconds of simulation can overpass the ability of GPs to simulate the system with a \textit{chirp+ramp} signal for all inputs. \textit{RNN 2} and \textit{RNN 1} require higher quantity of data of rich data to be effective in the simulation of AUV outside of the learning horizon in comparison to multi-output GPs.
\begin{table}[htbp]
  \centering
  \caption{RMSE results for all simulation, average and standard deviation}\centering
    \begin{tabular}{c|rrr}
    \multicolumn{1}{c}{\textbf{Simulation Number}} & \multicolumn{1}{c}{\textbf{GP}} & \multicolumn{1}{c}{\textbf{RNN1}} & \multicolumn{1}{c}{\textbf{RNN2}} \\
    \midrule
    1     & 1.85E-02 & 5.27E-02 & 6.34E-02 \\
    2     & 1.48E-02 & 1.76E-02 & 6.96E-02 \\
    3     & 1.69E-02 & 1.34E-02 & 3.83E-02 \\
    4     & 2.31E-02 & 5.12E-02 & 5.12E-02 \\
    5     & 2.88E-02 & 3.59E-02 & 4.07E-02 \\
    6     & 1.18E-02 & 1.68E-02 & 2.60E-02 \\
    7     & 1.46E-02 & 3.09E-02 & 2.70E-02 \\
    8     & 2.64E-02 & 2.02E-02 & 3.45E-02 \\
    \midrule
    \textbf{Average} & \textbf{1.94E-02} & \textbf{2.98E-02} & \textbf{4.38E-02} \\
    \midrule
    \textbf{Standard deviation} & 3.71E-05 & 2.43E-04 & 2.61E-04 \\
    \end{tabular}%
  \label{tab:rmse}%
\end{table}%
\begin{table}[htbp]
  \centering
  \caption{MAE results for all simulation, average and standard deviation}
    \begin{tabular}{c|rrr}
    \multicolumn{1}{c}{\textbf{Simulation Number}} & \multicolumn{1}{c}{\textbf{GP}} & \multicolumn{1}{c}{\textbf{RNN1}} & \multicolumn{1}{c}{\textbf{RNN2}} \\
    \midrule
    1     & 1.19E-02 & 2.04E-02 & 3.51E-02 \\
    2     & 9.51E-03 & 1.00E-02 & 4.02E-02 \\
    3     & 1.20E-02 & 8.87E-03 & 2.07E-02 \\
    4     & 1.59E-02 & 3.32E-02 & 3.03E-02 \\
    5     & 1.78E-02 & 2.08E-02 & 2.36E-02 \\
    6     & 8.43E-03 & 9.67E-03 & 1.48E-02 \\
    7     & 9.92E-03 & 1.82E-02 & 1.45E-02 \\
    8     & 2.21E-02 & 3.19E-01 & 1.60E-02 \\
    \midrule
    \textbf{Average} & \textbf{1.35E-02} & \textbf{5.50E-02} & \textbf{2.44E-02} \\
    \midrule
    \textbf{Standard deviation} & 2.27E-05 & 1.14E-02 & 9.58E-05 \\
    \end{tabular}%
  \label{tab:mae}%
\end{table}%

\begin{table}[htbp]
  \centering
  \caption{PRESS results for all simulation, average and standard deviation}
    \begin{tabular}{c|rrr}
    \multicolumn{1}{c}{\textbf{Simulation Number}}  & \multicolumn{1}{c}{\textbf{GP}} & \multicolumn{1}{c}{\textbf{RNN1}} & \multicolumn{1}{c}{\textbf{RNN2}} \\
    \midrule
    1     & 4.69E-01 & 7.89E+00 & 4.33E+00 \\
    2     & 2.61E-01 & 5.41E-01 & 9.96E+00 \\
    3     & 4.54E-01 & 2.14E-01 & 1.68E+00 \\
    4     & 9.58E-01 & 4.15E+00 & 2.91E+00 \\
    5     & 1.10E+00 & 2.03E+00 & 2.22E+00 \\
    6     & 1.54E-01 & 4.78E-01 & 8.60E-01 \\
    7     & 2.48E-01 & 2.12E+00 & 1.00E+00 \\
    8     & 1.18E+00 & 6.67E-01 & 1.50E+00 \\
    \midrule
    \textbf{Average} & \textbf{6.03E-01} & \textbf{2.26E+00} & \textbf{3.06E+00} \\
    \midrule
    \textbf{Standard deviation} & 1.70E-01 & 6.89E+00 & 9.05E+00 \\
    \end{tabular}%
  \label{tab:press}%
\end{table}%

\section[Conclusion]{Conclusion}
In this work, the use of multi-output GPs for the system identification of AUV dynamics was tested on a REMUS 100 AUV. It was demonstrated that the nonparametric multi-output GPs can model an AUV as well as RNN with the added value of a confidence measurement. In the simulations, GPs show a better ability than RNN to predict and simulate the behaviour of an AUV. In some cases, GPs performed better than RNN outside of the training horizons with the error between the GPs and the real system being relatively low as the convolution process is equivalent to represent the system through a differential equation. The GPs model obtained also has a smaller number of hyperparameters compared to the large number of coefficients of a mathematical model. The results of the sensitivity analysis show that multi-output GPs perform better that RNN with low quantities of data. RNN required a higher spectrum of data to be able to approximate the behaviour of the vehicle outside of the training horizon.

To improve further the capability of prediction of the system, more recent suggested techniques for GPs such as Recurrent GPs can be used. The simulation of GPs can also be improved if techniques such as Montercarlo and Taylor series can take advantage of the variance to increase the horizon of cover manoeuvres and the prediction accuracy. The next work will be devoted to the development of application for navigation and control using the obtained model. As the real world is a noisy environment that can be better described with Gaussian distributions, the use of GPs can show better performance in specific tasks such as navigation and control.

\bibliographystyle{unsrt}  
\bibliography{literature}  


\end{document}